\documentclass[lettersize,journal]{IEEEtran}
\IEEEoverridecommandlockouts

\def\BibTeX{{\rm B\kern-.05em{\sc i\kern-.025em b}\kern-.08em
    T\kern-.1667em\lower.7ex\hbox{E}\kern-.125emX}}
\usepackage[dvipsnames]{xcolor}
\usepackage{amsmath,amssymb,amsfonts,fancyvrb,url,soul}
\usepackage[T1]{fontenc}
\usepackage[utf8]{inputenc}
\usepackage[inline]{enumitem}

\usepackage{graphicx}
\usepackage{msc}
\usepackage{tikz}
\usetikzlibrary{matrix, arrows, automata, positioning, arrows.meta, shapes.misc}
\usepackage{colortbl}
\usepackage{pifont}
\usepackage{booktabs}
\usepackage{caption}

\usepackage{tabularray}
\UseTblrLibrary{booktabs}
\usepackage[numbers]{natbib} 
\usepackage{hyperref}
\usepackage{cleveref}

\usepackage{listings}
\usepackage{marvosym}
\usepackage{framed}
\usepackage{quoting}
\usepackage{subcaption}
\usepackage{multicol}
\usepackage{tcolorbox}
\tcbuselibrary{listings,skins}
\usepackage{float}
\usepackage{newfloat}
\usepackage[normalem]{ulem}
\usepackage{calc}
\usepackage{mathtools}
\usepackage{wasysym}

\definecolor{stackorangeback}{HTML}{fbdbc1}
\definecolor{stackorangeback1}{HTML}{e7700d}
\definecolor{stackblueback}{HTML}{d6e8fa}
\definecolor{stackblueback1}{HTML}{f2f8fd}
\definecolor{stackblueback2}{HTML}{1b75d0}
\definecolor{stackback}{HTML}{f6f6f6}
\definecolor{stackblue}{HTML}{013a65}
\definecolor{stackorange}{HTML}{b75301}
\definecolor{stackgreen}{HTML}{567a0d}
\definecolor{stackgreen1}{HTML}{18864b}
\definecolor{stackgreen2}{HTML}{0e6235}
\definecolor{stackred}{HTML}{9c2121}
\definecolor{stackred2}{HTML}{9c2121}

\definecolor{tcolboxbackground}{HTML}{ffffff}
\definecolor{rosepinedownpine}{HTML}{286983}
\definecolor{rosepinedownlove}{HTML}{b4637a}
\definecolor{rosepinedownfoam}{HTML}{56949f}
\definecolor{rosepinedownbase}{HTML}{faf4ed}
\definecolor{rosepinedownsurface}{HTML}{fffaf3}
\definecolor{rosepinedownhighlightlow}{HTML}{f4ede8}
\definecolor{rosepinedowngold}{HTML}{ea9d34}
\definecolor{rosepinedowntext}{HTML}{575279}
\definecolor{rosepinedownrose}{HTML}{d7827e}
\definecolor{rosepinedowniris}{HTML}{907aa9}
\definecolor{rosepinedownsubtle}{HTML}{797593}
\definecolor{rosepinedownmuted}{HTML}{9893a5}
\definecolor{rosepinedownoverlay}{HTML}{f2e9e1}

\hypersetup{colorlinks,linkcolor={black},citecolor={black},urlcolor={black}}

\newcommand{\newcheckmark}{\ding{51}}
\newcommand{\newcrossmark}{\ding{55}}

\newcommand{\chmark}[0]{{\newcheckmark}}
\newcommand{\xmark}{{\newcrossmark}}%

\newcommand{\msgcolor}[1]{\textcolor{Black}{#1}}

\DeclareRobustCommand{\xjoinrel}{\mathrel{\mkern-3.5mu}}

\newcommand{\snpgvm}[0]{\textsc{gvm}}
\newcommand{\snpgo}[0]{\textsc{go}}
\newcommand{\snpma}[0]{\textsc{ma}}
\newcommand{\snpfw}[0]{\textsc{fw}}
\newcommand{\amdkds}[0]{\textsc{kds}}

\newcommand{\tamarin}{\textsc{Tamarin}}

\makeatletter
\newcommand\BeraMonottfamily{%
  \def\fvm@Scale{0.85}
  \fontfamily{fvm}\selectfont
}
\makeatother

\makeatletter
\newcommand\AnonymousProttfamily{%
  \fontfamily{AnonymousPro}\selectfont
}
\makeatother

\newcommand{\dollar}{\mbox{\scriptsize\textdollar}}

\lstdefinestyle{mverb}{
  basicstyle=\ttfamily\footnotesize,
}

\lstdefinestyle{Tamarin}{
  upquote=true,
  basicstyle=\ttfamily,
  breaklines=true,
  morekeywords={rule,let,in,Fr,Out,pk,In,lemma,not,Ex,All,@,
    builtins,wrap,<,restriction,theory,begin,end,diff},
  otherkeywords={[,],-},
  mathescape,
}

\lstdefinestyle{Tamarin1}{
  frame=none,
  columns=flexible,
  basicstyle=\rmfamily,
  mathescape,
}

\lstdefinestyle{Tamarin2}{
  frame=none,
  numbers=left,
  stepnumber=1,
  showstringspaces=false,
  numbersep=4pt,
  numberstyle=\ttfamily\scriptsize\color[gray]{0.3},
  columns=flexible,
  basicstyle=\AnonymousProttfamily\footnotesize,
  morekeywords={rule,lemma,let,in,Fr,Out,pk,sign,verify, true,In,
    builtins,wrap,unwrap,senc,sdec,mac,kdf,<,restriction,theory,begin,end,diff},
  morestring=[b]',
  keywordstyle=\color{stackblue},
  stringstyle=\color{stackgreen},
  otherkeywords={[,],-},
  keywordstyle=[2]\color{stackblue},
  morekeywords=[2]{@,},
  mathescape,
  literate={dots}{{$\mathtt{\dots}$}}{1},
}

\lstdefinestyle{Tamarin3}{
  frame=none,
  columns=flexible,
  basicstyle=\AnonymousProttfamily\scriptsize{},
  mathescape,
  literate={dots}{{$\mathtt{\dots}$}}{1},
}

\lstdefinestyle{Tamarin4}{
  frame=none,
  numbers=left,
  stepnumber=1,
  showstringspaces=false,
  numbersep=4pt,
  numberstyle=\ttfamily\scriptsize\color[gray]{0.3},
  columns=flexible,
  basicstyle=\AnonymousProttfamily\footnotesize,
  mathescape,
  literate={dots}{{$\mathtt{\dots}$}}{1},
}

\newtcblisting{tcbTamarin2}{
      colback=stackback,
      top=-2mm,
      bottom=-2mm,
      left=0mm,
      right=0mm,
      boxrule=0pt,
      listing only,
      listing options={style=Tamarin4},
}

\newtcblisting{tcbTamarin2blue}{
      colback=stackback,
      top=-2mm,
      bottom=-2mm,
      left=0mm,
      right=0mm,
      boxrule=0pt,
      colback = blue,
      listing only,
      listing options={style=Tamarin4},
}

\newcommand{\mkwd}[1]{\color{black}\mathtt{#1}}

\newcommand{\smsn}[1]{$\color{darkgray}\tlstt{#1}$}

\newcommand{\tlst}[1]{\lstinline[style=Tamarin4,]|#1|}
\newcommand{\tlstt}[1]{\lstinline[style=Tamarin3,]|#1|}
\newcommand{\vlst}[1]{\lstinline[style=mverb,]|#1|}

\newcommand{\ltrm}[3]{\lstinline[style=Tamarin4,]|#1 $\mathrel{-}\xjoinrel\mathrel{[}$    #2 $\mathrel{]}\xjoinrel\rightarrow$ #3|}

 \colorlet{shadecolor}{gray!20}
\usepackage{lipsum}
\newenvironment{shadedquotation}
 {\begin{shaded*}\it
  \quoting[leftmargin=0pt, vskip=0pt]
 }
 {\endquoting
 \end{shaded*}
}

\DeclareFloatingEnvironment[fileext=lop]{lemma}

\newcommand\newsubcap[1]{\phantomcaption%
       \caption*{\normalsize\figurename~\thefigure\thesubfigure: #1}}

\newcommand\lemmanewsubcap[1]{\phantomcaption%
       \caption*{\normalsize\lemmaname~\thelemma\thesublemma: #1}}

\newcommand*\circled[1]{\tikz[baseline=(char.base)]{%
            \node[shape=circle,fill=darkgray!80,draw=darkgray!80,inner sep=1pt] (char) {\textcolor{white}{\small #1}};}}

\begin{document}

\title{Formal Security Analysis of the AMD SEV-SNP\\ Software Interface}

\pagestyle{plain}


\author{
  \IEEEauthorblockN{Petar Parad\v{z}ik},
  \IEEEauthorblockA{
	  \textit{University of Zagreb Faculty of Electrical Engineering and Computing},
    petar.paradzik@fer.hr\\
  }
  \and
  \IEEEauthorblockN{Ante Derek},
  \IEEEauthorblockA{
  \textit{University of Zagreb Faculty of Electrical Engineering and Computing},
    ante.derek@fer.hr\\}
  \and
  \IEEEauthorblockN{Marko Horvat},
  \IEEEauthorblockA{
    \textit{University of Zagreb Faculty of Science},
    marko.horvat@math.hr}
}

\maketitle

\begin{abstract}

  AMD Secure Encrypted Virtualization technologies enable confidential computing
  by protecting virtual machines from highly privileged software such as
  hypervisors.
  In this work, we develop the first, comprehensive symbolic model of the software interface of
  the latest SEV iteration called SEV Secure Nested Paging (SEV-SNP).
  Our model covers remote attestation, key derivation, page swap and live
  migration.
  We analyze the security of the software interface of SEV-SNP and formally prove that most critical secrecy, authentication, attestation and freshness properties do indeed hold in the model. 
  Furthermore, we find that the platform-agnostic nature of messages exchanged between SNP guests and the AMD Secure Processor firmware presents a potential weakness in the design.
  We show how this weakness leads to formal attacks on multiple security properties, including the partial compromise of attestation report integrity, and discuss possible impacts and mitigations.




\end{abstract}


\section{Introduction}





An increasing number of cloud providers are beginning to rely on trusted execution
environments (TEEs) to ensure the safety of user data while it is being processed by
applications and services on foreign platforms.
TEEs are tamper-resistant environments intended to provide the confidentiality and
integrity of user code and data in use, and isolate them from untrusted,
yet highly privileged software such as operating system kernels and
hypervisors. This makes TEEs suitable for
processing secrets on remote devices and platforms such as public clouds.
Modern TEEs also provide useful features such as remote attestation, a
mechanism by which a remote party can obtain evidence that a particular process
or virtual machine (VM) is running correctly with the expected configuration.

Hardware-based TEE solutions typically involve a security kernel executed on a main processor with
security extensions, or on a coprocessor. The security kernel is a piece of software hardwired on a chip (ARM TrustZone~\cite{arm}),
or it comes in the form of microcode
(Intel SGX~\cite{intel}) or firmware (AMD SEV~\cite{sev}) that resides in an isolated area within the processor, which
is not directly accessible or modifiable by external code. The software implements an interface
as an instruction set that can be used to establish a secure environment.

Multiple hardware vendors today offer competing TEE architectures with varying
levels of granularity. Intel provides Software Guard Extensions (SGX) to isolate
specific parts of applications (enclaves) and Trust Domain Extensions (TDX) to
protect the entire software stack of a VM. Similarly, ARM
offers TrustZone for isolating sensitive operations within a device and
Confidential Compute Architecture (CCA) for enhancing security in cloud and edge
computing environments~\cite{280904}.

Historically, the first of the hardware-based TEE solutions from AMD was named
Secure Virtualization (SEV)~\cite{sev}. Developed in 2016, SEV
was made available for the first generation of the AMD EPYC brand of x86-64
microprocessors, based on the Zen 1 microarchitecture codenamed Naples.
SEV enhanced VM security through VM
memory encryption and isolation. It also offered live VM migration
and launch attestation features. The latter represented a rather limited form of remote
attestation, allowing only the guest owner to attest the integrity of their guest VMs during the guest launch procedure.

A year later, AMD introduced SEV Encrypted State (SEV-ES)~\cite{sev-es}, making it available
for the second generation of AMD EPYC microprocessors
based on the Zen 2
microarchitecture codenamed Rome.  SEV-ES encrypts and protects the integrity of 
CPU registers that store information about the VM runtime before handing over 
control to a hypervisor, thereby preventing the hypervisor from reading
sensitive information of the guest as well as tampering with the control flow of the guest.


In this paper, we focus on AMD SEV Secure Nested Paging (SEV-SNP), which came out in 2020 and represents the
latest iteration of the AMD SEV technologies.
It is currently available for the third and fourth generation of the AMD EPYC
microprocessors, based on Zen 3 and Zen 4 microarchitectures codenamed Milan and
Genoa, respectively.
With the added
precaution of treating the hypervisor as fully untrusted, AMD SEV-SNP introduces memory integrity protection
as well as other security and usability enhancements to the existing SEV and SEV-ES
functionalities, including more flexibility of essential features such as live
migration and enabling remote attestation for third parties. AMD SEV-SNP is
used by popular cloud providers such as
Microsoft Azure~\cite{azure}, AWS~\cite{aws}, and Google
Cloud~\cite{googlecloud} to help provide Infrastructure as a Service.

There has been a substantial amount of analysis of the SEV technology
from both academia
and industry that, for the most part, focuses on its implementation.
However, to the best of our knowledge, there is no work
that formally analyzes the software interface of SEV-SNP.

AMD SEV-SNP is a complex system which supports a large number of guest policies and features, with most having multiple variants, and it uses an intricate key schedule. This makes it very difficult to ascertain whether an adversary
could launch an interaction attack by using the interface in an unexpected way.
The main research question that we would like to answer in this
work is the following:
\begin{shadedquotation}
Can an adversary use the AMD SEV-SNP software interface to force the system into an undesirable state?
\end{shadedquotation}






In this work, we employ the \textsc{Tamarin Prover} (\tamarin) protocol
verification tool to meticulously model and analyze the software interface of AMD
SEV-SNP.  The model encompasses all key features and captures many
subtle behaviors of SEV-SNP. We specify and analyze nearly a hundred properties
and find several formal attacks.
Specifically, our contributions are as follows:
\begin{itemize}
  \item We develop a formal model of the AMD SEV-SNP software interface.  The
        model is close to being fully comprehensive; it covers protected guest launch, remote
        attestation, key derivation, page swap and live migration.
        To the best of our
        knowledge, there exists no prior formal model of the SEV-SNP software interface.
  \item We give formal definitions of, and automated proofs for, critical secrecy, authentication,
        attestation, and freshness properties. We show that for the most part, AMD SEV-SNP software interface does indeed meet the desired security goals.
        This includes the proof of correct
        stream cipher usage.
  \item We show that the platform-agnostic nature of
        SNP-protected guest messages
        leads to formal attacks on several
        authentication and attestation properties,
        including the compromise of attestation report integrity.
  \item We discuss the challenges of fully mitigating the discovered weaknesses
        while preserving the seamless guest VM migration feature.  Additionally, we show
        that the vague specification of migration agents is consistent with
        a scenario where a cloud
        provider may trick a third party into incorrectly believing that an
        SNP-protected guest is running on a platform with secure, up-to-date firmware,
        and suggest possible mitigations.
\end{itemize}

\noindent\emph{Paper Outline}. We first give the necessary background related to AMD SEV-SNP and the \tamarin{} tool in Section~\ref{sec:background}. Next, we explain the minutiae of our formal model of AMD SEV-SNP in Section~\ref{sec:model}. We provide the details of our analysis in Section~\ref{sec:analysis} and thoroughly discuss both the positive and negative results in Section~\ref{sec:results}; we also suggest possible countermeasures. Finally, we give an overview of the related work in Section~\ref{sec:relatedwork} and conclude in Section~\ref{sec:conclusion}.

\section{Background}\label{sec:background}



The core ideas behind SEV-SNP were outlined in a white paper published in January
2020~\cite{sevsnp:whitepaper}, and initially
specified in April 2020. In June 2022, Linux 5.19-rc1 was released with partial SEV-SNP
support~\cite{sevsnp:linux}, and in August 2023, AMD made the SEV-SNP Genoa
firmware source code publicly available on GitHub~\cite{sevsnp:github}.
The current revision of the specification was released in September
2023~\cite{sevsnp:spec}.

SEV-SNP provides
a novel application binary interface (ABI) intended for a hypervisor to
bootstrap and manage virtual machines within a secure environment.
It extends SEV technologies by enhancing memory integrity protection, attestation,
and virtualization capabilities. In particular, it
\begin{itemize}
  \item tracks the owner of each page of memory;
  \item utilizes Trusted Computing Base (TCB) versioning to guarantee
        that guests run under up-to-date firmware;
  \item adds a versatile remote attestation mechanism where a third party may
        establish trust in a guest during runtime of the guest; 
  \item supports generating guest key material from different sources;
  \item enhances the flexibility of live virtual machine migration by introducing
        an entity called a \emph{Migration Agent};
  \item provides several additional features such as secure nested virtualization.
\end{itemize}



SEV-SNP prevents a hypervisor from replacing VM memory with an old copy (\emph{replay attack})
or mapping two different guest physical addresses to a single system
physical address or DRAM page (\emph{memory aliasing attack}).
It does so by utilizing a data structure called
Reverse Map Table (RMP) and a page validation mechanism to track page
ownership and enforce proper page access control.

In previous SEV instances, a hypervisor was assumed not to be malicious, but potentially buggy. Like before, from the
perspective of a single SNP-protected guest, other
VMs---whether non-SNP (legacy) or SNP-protected---are also regarded as untrusted
entities. However, unlike before, SEV-SNP treats
the hypervisor as fully untrusted, capable of tampering with page tables, injecting
arbitrary events, and providing false system information. 

The TCB of SEV-SNP comprises the AMD System on Chip,
which includes the AMD Security Processor (AMD-SP), and the
software running on top of
it, which currently comprises the microcode, bootloader, operating system, and
the SNP firmware which implements the SNP ABI.
Each TCB software component can be upgraded, and its security version
numbers are included in the \vlst{TCB_VERSION} structure that is associated
with each SNP VM image.

SEV-SNP introduces an interface that the SNP-protected guest may utilize to
request services from the firmware during runtime.  This enables the guest
to obtain---via so-called \emph{guest messages}---attestation reports and derived
keys, for instance. These messages are encrypted and integrity protected
by the Virtual Machine Platform Communication Key (\vlst{vmpck}).
The firmware generates and installs this key into both the guest context that it
maintains and the guest memory pages during the guest launch.


In SEV and SEV-ES, the attestation procedure was confined to the guest launch,
limiting its use to a singular entity---the guest owner.
SEV-SNP introduces a more versatile approach by replacing launch attestation
with remote attestation, enabling any third party to acquire an attestation
report and establish trust in a guest at any given moment.



The attestation report comprises various information about the guest, such as
the guest policy, image digest (\emph{measurement}), the platform (chip
identifier) it operates on, and TCB version.
It is signed using a private ECDSA P-384 key that is either a
Versioned Chip Endorsement Key (\vlst{VCEK})---a machine-specific key derived
from chip-unique secrets and a digest of \vlst{TCB_VERSION}---or a Versioned
Loaded Endorsement Key (\vlst{VLEK}), which is a cloud provider-specific key derived from a seed
maintained by the AMD Key Derivation Service (KDS).  The inclusion of the TCB
version allows a third party to reject the signature if it originated from an unpatched AMD-SP.



In order to further demonstrate the authenticity of the attestation report signature,
the \vlst{VCEK}/\vlst{VLEK} is verified against the AMD Signing Key (\vlst{ASK}), which is further
verified against the AMD Root Key (\vlst{ARK}), both of which are 4096-bit RSA keys.

SEV-SNP introduces a robust key derivation mechanism. It enables an
SNP-protected guest to instruct the firmware to derive a key
rooted in either \vlst{VCEK}, \vlst{VLEK}, or a Virtual Machine Root Key
(\vlst{VMRK}).  Moreover, the guest has the option to request adding further data, such as its launch digest and TCB version, into the key derivation process.
Such keys may be used for data sealing and other purposes.

Similar to previous SEV technologies, SEV-SNP offers an interface for secure
swapping wherein the guest may be saved to disk and later resumed. The
firmware ensures confidentiality and authenticity of the guest memory pages
by utilizing an Offline Encryption Key (\vlst{OEK}).
The swapping plays a crucial role in live migration.

Live migration is a mechanism by which guests may be seamlessly and securely transferred to another
physical system, without having to shut themselves down first.
Whereas in SEV and SEV-ES
the firmware on the source machine was responsible for authenticating
the firmware on the destination machine prior to guest context transfer, in
SEV-SNP this task is facilitated by an entity called a Migration
Agent.

A Migration Agent is an SNP-protected guest that is responsible for enforcing
guest migration policies and providing the firmware with guest-unique
secrets (\vlst{VMRK}).  Whereas a single guest may be associated with at most one
migration agent, a single migration agent may be associated with and manage
multiple guests concurrently.
This association is indicated in each attestation report.
SEV-SNP does not specify the behavior of migration agents nor the
manner by which the guest context is securely transferred over the network.

\subsection*{Tamarin Prover}

\textsc{Tamarin Prover} (\tamarin{})~\cite{tamarin} is an automated symbolic verification tool for security protocols. It is based on multiset rewriting;
more precisely,
its semantics comprises a labeled transition system whose states are multisets
of \emph{facts}, and the transitions between them are specified by \emph{rules}
which prescribe the behavior of protocol participants and the adversary.
Each rule has a
\emph{left-hand side}
(facts that must be available in the current global state for the
rule to execute), \emph{actions} (labels which are logged in the trace and used to express the desired
security properties), and a \emph{right-hand side} (facts that will be added to
the state).

The left-hand and right-hand side contain multisets of facts, each fact being either \emph{linear} or \emph{persistent} (the latter are prefixed with an exclamation point). Facts can be \emph{produced} (when executing a rule with the fact on the right-hand side) and \emph{consumed} (if it is both linear and on the left-hand side). While there is no bound on the number of times a fact can be produced, linear facts model limited resources that can only be consumed as many times as they are produced, and persistent facts model unlimited resources, which (once produced) can be consumed any number of times.

\emph{Fresh} variables, denoted by the prefix ``\tlst{\~}'' (or suffix
``\tlst{:fresh}''), indicate freshly generated names. They are suitable for modelling
randomly generated values such as keys and thread identifiers. The built-in fact
\tlst{Fr($\mathtt{\cdot}$)} can be utilized to generate such names.
\emph{Public} variables, identified by the prefix ``\tlst{$\dollar$}'' (or
suffix ``\tlst{:pub}''), are used to represent publicly known names such as
agent identities and group generators. Additionally, \emph{temporal} variables,
prefixed with ``\tlst{#}'', signify timepoints.





The \tamarin{} \emph{builtins} include equational theories for Diffie-Hellman
operations, (a)symmetric encryption, digital signatures, and hashing.
The theory \emph{natural-numbers}
can be used to
model monotonic counters.
Additionally, \tamarin{} supports user-defined function symbols and equational theories. For example, in
our model, two ternary function symbols, \tlst{wrap} and \tlst{unwrap} are
defined, along with the equation
\begin{tcbTamarin2}
  unwrap(wrap(msg, nonce, key), nonce, key) = msg
\end{tcbTamarin2}
\noindent which models a symmetric cipher with a nonce as an initialization vector; incrementing the nonce is modelled by using the built-in \emph{natural-numbers} theory.

Consider the following rule (taken from our SEV-SNP
model and simplified) that enables an SNP-protected guest to
request an attestation report.


\begin{tcbTamarin2}
$[$ StateGvm('RUNNING', $\dollar$image, ~key, %nonce, ~ptr)
, In(rd), Fr(~newPtr) $]$
$\mathrel{-}\xjoinrel\mathrel{[}$ ReportRequest($\dollar$image, rd) $\mathrel{]}\xjoinrel\rightarrow$
$[$ Out(wrap(< 'MSG_REPORT_REQ',   rd >, %nonce, key))
, StateGvm('WAIT', $\dollar$image, ~key, %nonce, ~newPtr) $]$
\end{tcbTamarin2}

The rule requires that a
guest is in the \tlst{RUNNING} state before it can be executed: it consumes the
linear fact \tlst{StateGvm('RUNNING', $\dollar$image, ~key, \%nonce, ~ptr)} from the global
state, receives an arbitrary message \tlst{rd} (which will be included in the
attestation report) from the network using the \tlst{In($\mathtt{\cdot}$)} fact,
and generates a new pointer \tlst{newPtr} to its updated state using the
\tlst{Fr($\mathtt{\cdot}$)} fact.

Subsequently, the guest constructs a plaintext, which is an ordered pair
\tlst{<'MSG_REPORT_REQ', rd >}
comprising the message tag and received data,
encrypts it with a symmetric \tlst{key} using a \tlst{\%nonce}
and obtains the ciphertext
\tlst{wrap(<'MSG_REPORT_REQ', rd >, key, \%nonce)},
sends the ciphertext to the network using the \tlst{Out($\mathtt{\cdot}$)} fact,
and a new linear fact
\tlst{StateGvm('WAIT', $\dollar$image, \~key, \%nonce, \~newptr)} is produced. This fact can now be
consumed by a rule wherein the guest receives the attestation report.


\tamarin{} assumes a Dolev-Yao adversary who relays all messages between protocol participants.
More precisely, the \tlst{KU} and \tlst{KD} facts represent the adversary knowledge
and its ability to receive messages from the network and send messages to the network, respectively, by
using
\lstinline[style=Tamarin4,]|Out(x)$\mathrel{\mathrel{-}\xjoinrel\mathrel{[}\mathrel{]}\xjoinrel\rightarrow}$KD(x)| and
\lstinline[style=Tamarin4,]|KU(x)$\mathrel{-}\xjoinrel\mathrel{[}$ K(x) $\mathrel{]}\xjoinrel\rightarrow$In(x)|
communication rules (as with all actions, the \tlst{K} action is part of the \tamarin{} property specification language and can be used to express that the adversary necessarily knows the term it sends to the network). The adversary can try to deconstruct messages 
it received (constructing any keys needed) in order to gain knowledge of additional terms
(e.g. decrypt a ciphertext with the help of the rule
\lstinline[style=Tamarin4,]|KD(wrap(x,y)), KU(y)$\mathrel{-}\xjoinrel\mathrel{[}$ $\mathrel{]}\xjoinrel\rightarrow$KD(x)|)
and then switch via
\lstinline[style=Tamarin4,]|KD(x)$\mathrel{-}\xjoinrel\mathrel{[\ ]}\xjoinrel\rightarrow$KU(x)|
to constructing new
messages by applying cryptographic operations on known messages (e.g. applying a hash function can be done with the rule
\lstinline[style=Tamarin4,]|KU(x)$\mathrel{-}\xjoinrel\mathrel{[\ ]}\xjoinrel\rightarrow$KU(h(x))|).




The trace properties
of a protocol encoded as a set of multiset-rewriting rules
are specified as temporal first-order formulas (\emph{lemmas}) over the rule labels.
For example, the following secrecy property states that,
for all SNP firmware threads (\tlst{fwId}), guest VM contexts (\tlst{gvmCtx}), VM Platform Communication
Keys (\tlst{vmpck}), and timepoints \tlst{#i} and \tlst{#j},
if a thread with the identifier \tlst{fwId} generates \tlst{vmpck} and
installs it in \tlst{gvmCtx} at some timepoint \tlst{#i}.
and if the adversary knows \tlst{vmpck} at \tlst{#j},
then it necessarily holds that there exists some timepoint \tlst{#k}
such that the adversary corrupted \tlst{vmpck} at  \tlst{#k}
and \tlst{#k} precedes \tlst{#j}:


\begin{tcbTamarin2}
lemma SecMaVmpckIsSecret:
$\mkwd{\forall}$ fwId gvmCtx vmpck #i #j.
  Install('MA', fwId, gvmCtx, vmpck)@i $\mkwd{\land}$ KU(vmpck)@j
  $\mkwd{\Rightarrow}$ $\mkwd{\exists}$ #k. Corrupt(vmpck)@k $\mkwd{\land}$ (#k < #j)
\end{tcbTamarin2}

\tamarin{} proves trace properties by \emph{falsification}. In order to prove
that a property is true, \tamarin{} tries to find a counterexample execution---an
alternating sequence of states and transitions that satisfies
the negation of the property in question.
If \tamarin{} halts the analysis and succeeds, then the resulting execution represents an attack;
if \tamarin{} halts the analysis and fails, then it provides a proof of the property in the form of a tree.
Lastly, \tamarin{} may not terminate as verifying security properties is in general undecidable~\cite{mitchell1999undecidability}.
There are several ways to avoid non-termination and to improve verification time;
we customize the ranking of proof goals by writing scripts called \emph{(proof) oracles} and
use supporting lemmas to prove other lemmas.


Finally, \emph{restrictions} can be used to filter out traces that need not be
considered during the security analysis or to enforce certain behavior such as
verification of signatures or branching. One such restriction in our model
states that every memory pointer can only be read before it is released.


\begin{tcbTamarin2}
$\mkwd{\forall}$ ptr #i #j. Read(ptr)@i $\mkwd{\land}$ Free(ptr)@j $\mkwd{\Rightarrow}$ #i < #j
\end{tcbTamarin2}

\section{Formal model of AMD SEV-SNP}\label{sec:model}


Our goal is to model and analyze the software interface of AMD SEV-SNP. We aim
to capture all of its principal features, including remote attestation, key
derivation, page swapping and live migration, while ignoring the low-level details such as
the memory encryption and RMP structure.
We consider a powerful Dolev-Yao adversary that has full control over
the communication network in an idealized cryptography setting.
Before we delve into the intricacies of our model, we briefly explain the methodology we use.



We selected \textsc{Tamarin Prover} as widely accepted verification tool for
security protocols which supports loops, branching and mutable global state.
These features are essential to properly model the behavior of SNP-protected
guests, as they can be executed indefinitely, launched with different policies
and suspended temporarily due to swapping.

To facilitate development, we use the \vlst{m4} general-purpose macro processor.  This
allows for efficient prototyping (e.g., by disabling certain precomputations)
and provides the flexibility to extract multiple variants of the model. Some
variants, such as the one obtained with the flag \tlst{-DIGNORE_ROOT_MD_ENTRY},
are intentionally not aligned with the specification; we use them
\emph{sanity checks}, i.e., to confirm the necessity of certain integrity checks.


Our model is based on the specification \emph{SEV Secure Nested Paging Firmware
ABI Specification}, revision 1.55, published in September
2023~\cite{sevsnp:spec}. AMD upstreamed SEV-SNP guest Linux support~\cite{sevsnp-guest:github}
and recently published the source code of the SEV-SNP
Genoa firmware on GitHub~\cite{sevsnp:github}.  In scenarios where the
specification was not clear enough, we referred to the available implementations.



Although AMD has published in the SEV-SNP white paper~\cite{sevsnp:whitepaper} a
set of security threats (properties) that it addresses, we do not consider them
in our analysis. This is because they account for low-level behavior, such as
memory aliasing, rather then the specific manner in which the interface is
utilized.  Since no security properties are defined in the specification either,
we defined our own set of properties.

Although we believe that the formal attacks we have identified are consistent
with both the specification and its implementation, we have not yet been able to
execute them on actual hardware.  This is due to the fact that, as of the time
of writing, the SEV-SNP live migration feature is not fully supported in
open-source tools like QEMU/KVM.

Our model, complete with proofs and documentation, is available on
Gitlab~\cite{themodel:gitlab}.

\subsection{Entities}

The model can be roughly described in terms of five, somewhat simplified,
interacting state machines. In particular, we have:

\vspace{5pt}
\begin{itemize}\setlength\itemsep{0.5em}
  \item a state machine that describes the behavior of the \textsc{SNP}-protected Guest (\textsc{gvm})
        as depicted in~\Cref{fig:GVMSM};
  \item a state machine that describes the behavior of the Guest Owner (\textsc{go})
        as depicted in~\Cref{fig:GOSM};
  \item a state machine that describes the behavior of the Migration Agent (\textsc{ma})
        as depicted in~\Cref{fig:MASM};
  \item state machines that describe the behavior of the SNP Firmware (\textsc{fw}):
         one that launches \textsc{gvm}s, and one that responds to \textsc{gvm} requests,
         as depicted in~\Cref{fig:ASPFWLaunchSM} and~\Cref{fig:ASPFWCommSM}, respectively.
\end{itemize}
\vspace{5pt}




In our model, multiple entities can interact indefinitely, in an arbitrary
order, with the SEV-SNP software interface.
To support this interaction, we allow an unbounded execution of each \emph{thread},
regardless of the program it executes (i.e., the entity it belongs to).
For example, a single \snpma{} thread can manage, i.e.\ be associated with, an
unbounded number of \snpgvm{}s.
If a thread is capable of executing multiple tasks, it can execute
them in any order.  For example, a \snpgvm{} thread can execute any sequence of
attestation report and key derivation requests.
In some cases, the order in which
calls are made may result in substantially different traces as they may update a shared
state. For example, if the \snpgvm{} gets swapped out immediately after
it receives a derived key, its encrypted page (state) will contain the key.




\subsection{Key Distribution Service}

The rules \tlst{KDSCreateARK} and \tlst{KDSCreateASK} model a part of the AMD
Key Distribution Service (\amdkds). This includes the generation of the AMD Root Key, denoted as
\tlst{privArk}, and the AMD Signing Key, denoted as \tlst{privAsk}. Each long-term key (including \tlst{privVcek} which is discussed
next) is available via the fact \tlst{!LTK} and
certified by the next key in the hierarchy,
except for \tlst{privArk}, which represents the root of trust and is
self-signed.
For the purpose of producing and verifying digital signatures, we
employ the built-in message theory \tlst{signing}, which exports function symbols
\tlst{sign}, \tlst{verify}, \tlst{pk}, \tlst{true}; they are related by the equation
\tlst{verify(sign(m,sk),m,pk(sk)) = true}.



We publish the certificate of each
long-term key to the network using \tlst{Out} facts,
and make the root certificate available to a guest owner (\snpgo{}) via \tlst{!Cert(dots)}
fact. Furthermore, we allow the adversary to compromise the \amdkds{} and extract the
keys using the \tlst{RevealARK} and \tlst{RevealASK} rules.

\subsection{Platform}






\begin{figure}
\begin{tcbTamarin2}
rule ChipCreate[color=colorKDS]:
let fwVcek = kdf('VCEK', cek, $\dollar$fwTcbVersion)
     fwPubVcek = pk(fwVcek)
     data = <'VCEK', askId, $\dollar$chipId, fwPubVcek>
     cert = <data, sign(data, privAsk)>
in
$\mathtt{[}$ !LTK('ASK', askId, privAsk), Fr(cek) $\mathtt{]}$
$\mathtt{\mathrel{-}\xjoinrel\mathrel{[}}$ Uniq(<'VCEK', $\dollar$chipId>)
 , ChipGenerateCek($\dollar$chipId, cek)
 , FwDeriveVcek($\dollar$chipId, fwVcek) $\mathtt{\mathrel{]}\xjoinrel\rightarrow}$
$\mathtt{[}$ !LTK('VCEK', $\dollar$chipId, fwVcek), CEK($\dollar$chipId, cek)
, Out(cert) $\mathtt{]}$
\end{tcbTamarin2}
\caption{Platform initialization}
\label{fig:chipcreate}
\end{figure}

We model the creation, initialization (\vlst{SNP_INIT}) and configuration
(\vlst{SNP_CONFIG}) of the platform by using the \tlst{ChipCreate}
and \tlst{PlCreate} rules.
This former rule (\Cref{fig:chipcreate}) binds a firmware to a specific chip, which
is denoted by \tlst{chipId}: it uses a freshly generated, chip-unique, long-term secret \tlst{cek}, the
value of a public variable \tlst{$\dollar$fwTcbVersion} (the TCB version of the firmware image) and
a key derivation function \tlst{kdf} to derive an
attestation signing key, \tlst{privVcek}. We model side-channel attacks on its confidentiality with the help of the \tlst{ExtractCek} rule.

The latter rule creates a pair of associated migration agent images, and bind each
to a particular platform.

Note that firmware updates (\vlst{DOWNLOAD_FIRMWARE}) are not supported; we represent all of \vlst{CurrentTcb},
\vlst{ReportedTcb} and \vlst{LaunchTcb} by \tlst{$\dollar$fwTcbVersion}, whose value (once initialized)
persists throughout the lifetime of a chip, and consequently the lifetime of \tlst{privVcek}.

The per-chip uniqueness of
\tlst{privVcek} is enforced by applying the restriction \tlst{Unique}.

\begin{tcbTamarin2}
restriction Unique:
$\mkwd{\forall}$ x #i #j. Uniq(x)@i $\mkwd{\land}$ Uniq(x)@j $\mkwd{\Rightarrow}$ #i = #j
\end{tcbTamarin2}

It requires that the \tlst{Uniq} action shown above be injective, so the restriction ensures that every instance of the \tlst{PLCreate} rule yields a distinct
\tlst{$\dollar$chipId}.

The fact
\tlst{!LTK('VCEK', dots)} may be used to start any number
of \snpfw{} threads, all running on the same chip; the details will be given
in the Initialization subsection.






\subsection{Measurement}
\label{sec:measurement}

A guest image measurement (digest) is computed by the guest owner prior to image deployment
and afterwards by the firmware during guest launch.
In the latter case, the firmware constructs the measurement by initializing a load digest via an \vlst{SNP_LAUNCH_START} call and subsequently updating it with \vlst{SNP_LAUNCH_UPDATE} calls. Each update saves a hash of the \vlst{PAGE_INFO} structure, which transitively binds the contents of each individual page to the digest.

When \vlst{SNP_LAUNCH_FINISH} is called, assuming that \vlst{ID_BLOCK_EN} is set, the firmware checks whether the computed measurement matches the one specified by the guest owner; if it does not, the firmware refuses to launch a guest and returns \vlst{BAD_MEASUREMENT}.

Note the dual nature of a guest image as both a program, which is intended to be executed on a virtual machine, and data that can be measured. In order for us to directly model both the dynamic and static nature of the image, the modelling language would need to support metaprogramming
features that would allow us to treat programs as data. To the best of our
knowledge, none of the current security protocol
verification tools offers such support.

We employ the following approach: to model an image
as data, we use a public variable \tlst{$\dollar$image}, and constants
\tlst{'5XPYKIAXFS06O'} and \tlst{'3A9B8C7D1E2F'} to
represent specific images assigned to a guest (\snpgvm{})
and a migration agent (\snpma{}), respectively.
To model an image as a program, we use a set of multiset-rewriting rules
and prefix each rule belonging to the respective images 
with \tlst{GVM} and \tlst{MA}.
By doing so we
assign a predefined load digest
to each image we consider (e.g.,
\tlst{h(< 'VM_IMAGE', '5XPYKIAXFS06O' >)}).

Whenever a \snpfw{} launches a \snpgvm{} or
\snpma{}, the behavior of the launched virtual machine is
governed by a set of multiset-rewriting rules which is fixed in advance.
While tampering with the images is not supported, we allow the adversary to compromise
the secrets contained in certain images.




\begin{figure*}
  \centering
  \begin{subfigure}{.6\textwidth}
  \begin{tikzpicture}[shorten >=1pt,node distance=2cm,auto]
    \tikzset{every node/.style={font=\scriptsize{}}}

    \tikzstyle{gvm state} = [
    draw,
    thick,
    row sep=1mm,
    text width = 3cm,
    rounded corners,
    ]

    \tikzstyle{gvm state name} = [
    text centered,
    inner sep = 0pt,
    outer sep = 0pt,
    text=darkgray,
    ]
    \tikzstyle{gvm state desc} = [
    inner sep = 1pt,
    ]

    \matrix [draw, rounded corners,thick, inner sep = 1pt, text width = 1cm, fill=gray, ] (gvm) at (2,1)
    {
      \node[text centered, thick, text=white] {$\mathsf{GVM}$}; \\
    };

    \matrix [gvm state,] (idle) [below = 0.4cm of gvm]
    {
      \node[gvm state name] {\smsn{IDLE}}; \\
      \node[gvm state desc] {
      Delete key (if derived)
      };\\
    };

    \matrix [gvm state, text width = 3.6cm,] (key-request) [below left = 0.4cm and -1.4cm of idle]
    {
      \node[gvm state name] {\smsn{RUNNING}};\\
      \node[gvm state desc] {
          Select root \tlstt{vmrk}\\
          Mix \tlstt{vmpl + hostData + idKey}\\
          Write \msgcolor{\tlstt{MSG\_KEY\_REQ}}\\
      };\\
    };

    \matrix [gvm state,] (key-response) [below = 0.4cm of key-request]
    {
      \node[gvm state name,] {\smsn{KEY\_REQUEST}}; \\
      \node[gvm state desc] {
        Save \tlstt{key} in memory
      }; \\
    };

    \matrix [gvm state, text width = 3.6cm] (report-request) [below right = 0.4cm and -1.4cm of idle]
    {
      \node[gvm state name] {\smsn{RUNNING}}; \\
      \node[gvm state desc] {
        Set \tlstt{reportData = nonce}\\
        Write \msgcolor{\tlstt{MSG\_REPORT\_REQ}} \\
      }; \\
    };

    \matrix [gvm state] (report-response) [below = 0.4cm of report-request]
    {
      \node[gvm state name] {\smsn{REPORT\_REQUEST}};\\
      \node[gvm state desc] {Send \msgcolor{\tlstt{REPORT\_CERT}} }; \\
    };

    \draw [-Latex,] (gvm.south) --  (idle.north);

    \path (idle.south west) -- (idle.south) coordinate[pos=0.5] (idle-mid-sw);
    \draw [-Latex,] (idle-mid-sw) --  (idle-mid-sw|- key-request.north);

    \draw [-Latex,] (key-request.south) -- node[left] {Read \msgcolor{\tlstt{MSG\_KEY\_RSP}}} (key-response.north);

    \path (idle.south east) -- (idle.south) coordinate[pos=0.5] (idle-mid-se);
    \draw [-Latex,] (idle-mid-se) -- node[right] {Receive \msgcolor{\tlstt{REPORT\_DATA}} (\tlstt{GO})} (idle-mid-se|- report-request.north);

    \draw [-Latex,] (report-request.south) -- node[right] {Read \msgcolor{\tlstt{MSG\_REPORT\_RSP}}} (report-response.north);

    \path (idle.north west) -- (idle.north) coordinate[pos=0.5] (idle-mid-nw);
    \draw [-Latex,] (key-response.south)
    |- ([yshift=-.2cm,xshift=-0.7cm]key-request.south west |- key-response.south west)
    |- ([yshift=0.3cm]idle.north west -| idle-mid-nw)
    -- (idle-mid-nw)
    ;

    \path (idle.north east) -- (idle.north) coordinate[pos=0.5] (idle-mid-ne);
    \draw [-Latex,] (report-response.south)
    |- ([yshift=-.2cm,xshift=1.2cm]report-request.south east |- report-response.south east)
    |- ([yshift=0.3cm]idle.north east -| idle-mid-ne)
    -- (idle-mid-ne)
    ;

  \end{tikzpicture}

  \newsubcap{SNP-protected Guest State Machine}
  \label{fig:GVMSM}
  \end{subfigure}%
  \begin{subfigure}{.3\textwidth}
  %
  \begin{tikzpicture}[shorten >=1pt,node distance=2cm,auto]
    \tikzset{every node/.style={font=\scriptsize{}}}

    \tikzstyle{gvm state} = [
    draw,
    thick,
    row sep=1mm,
    text width = 3cm,
    rounded corners,
    ]

    \tikzstyle{gvm state name} = [
    text centered,
    inner sep = 0pt,
    outer sep = 0pt,
    text=darkgray,
    ]
    \tikzstyle{gvm state desc} = [
    inner sep = 1pt,
    ]

    \matrix [%
    draw,
    rounded corners,
    thick,
    inner sep = 1pt,
    text width = 1cm,
    fill=gray,
    ] (GO) at (2,1)
    {
      \node[text centered, thick, text=white] {$\mathsf{GO}$}; \\
    };

    \matrix [gvm state, text width = 4.3cm] (GOInit) [below = 0.4cm of GO]
    {
      \node[gvm state name] {\smsn{INITIALIZE}}; \\
      \node[gvm state desc] {
          Set migration policy \tlstt{gvmMigPolicy} \\
          Calculate launch digest\\
          Prepare \tlstt{image}, sign \tlstt{idBlock}, set \tlstt{idKey}\\
          Send \msgcolor{\tlstt{DEPLOY\_REQ}}
      };\\
    };

    \matrix [gvm state] (GOReportReq) [below = 0.4cm of GOInit]
    {
      \node[gvm state name] {\smsn{REPORT\_REQUEST}};\\
      \node[gvm state desc] {
          Generate \tlstt{nonce}\\
          Send \msgcolor{\tlstt{REPORT\_DATA}}\\
      };\\
    };

    \matrix [gvm state] (GOVerifyReport) [below = 0.4cm of GOReportReq]
    {
      \node[gvm state name] {\smsn{REPORT\_VERIFY}};\\
      \node[gvm state desc] {
        Verify certificate with \tlstt{VCEK}
      };\\
    };

    \draw [-Latex,] (GO.south) --  (GOInit.north);
    \draw [-Latex,] (GOInit.south) --  (GOReportReq.north);
    \draw [-Latex,] (GOReportReq.south) -- node[left] {Receive \msgcolor{\tlstt{REPORT\_CERT}}}  (GOVerifyReport.north);

    \path (GOReportReq.north east) -- (GOReportReq.north) coordinate[pos=0.5] (go-report-req-mid-ne);
    \draw [-Latex,] (GOVerifyReport.south) |- ++(2.0, -0.2)
    |- ([yshift=0.3cm]go-report-req-mid-ne |- GOReportReq.north)
    -- (go-report-req-mid-ne)
    ;

  \end{tikzpicture}
  \newsubcap{Guest Owner State Machine}
  \label{fig:GOSM}
  \end{subfigure}

\end{figure*}

\subsection{Initialization}

Upon startup and before initiating any primary guest launch, \snpfw{} spawns a
Migration Agent (\snpma{}) background thread capable of managing guests
associated with it.  The \snpma{} thread should be able to migrate only those
guests associated with it during launch, where the association with a particular
guest depends on the migration policy of that guest.  However, because it is
constantly running in the background, the \snpma{} thread may attempt to initiate migration of
any guest at any time and potentially violate attestation report integrity if
the guest migration policy is not properly enforced.  We prove that such
violations are not possible in our model
(\tlst{AttReportIntegrityMd}).


By allowing a single \snpma{} thread to concurrently manage multiple guests, we address scenarios where the
adversary may replay the \snpma{} thread messages to multiple guests
associated with the thread. For instance, the adversary might be able to reuse the
\tlst{MSG_VMRK_REQ} message and install the same Virtual Machine Root Key
(\tlst{vmrk}) in more than one guest context, thus making
each of the guests later derive the same key rooted in
\tlst{vmrk}; protecting against such behavior is important because the derived key may be used e.g.\ for key sealing. We prove that the described
behavior is not possible in our model (\tlst{FreshKeyDerFromFwVmrkIsGvmUnique}).

The \tlst{FwLaunchMa} rule, as shown in~\Cref{fig:fwlaunchesma}, is an abstraction of all the \vlst{SNP_LAUNCH} commands. It models the launch of an \snpma{} thread by \snpfw{}.
The \snpma{} thread context \tlst{maCtx} is initialized with
several fresh values:
a firmware thread identifier \tlst{fwId}, which uniquely identifies a \snpfw{} thread;
an attestation report identifier \tlst{reportId}, which binds an attestation
report to a specific guest instance;
a secret Virtual Machine Platform Communication Key \tlst{vmpck} and
the corresponding message counter \tlst{nonce},
which are used to establish a secure communication channel between \snpgvm{} and
\snpfw{}; and
a pointer \tlst{maStPtr} to \tlst{maCtx} which is freshly generated upon each \tlst{maCtx} update.

\begin{figure}
\begin{tcbTamarin2}
rule FwLaunchMa:
let
  nonce = %1
  ma = <image, ~maId>
  image = '3A9B8C7D1E2F'
  ld = h(<'MA_IMAGE', image>)
  maCtx = <ld, ~vmpck, %nonce, ~reportId, assocPl dots>
  dots
in
$[$ Platform(plNum, chipId, fwVcek, ma, assocPl)
, Fr(~vmpck), Fr(~reportId), Fr(~fwId), Fr(~maStPtr) $]$
$\mathrel{-}\xjoinrel\mathrel{[}\dots \mathrel{]}\xjoinrel\rightarrow$
$[$ !StateFwMa('RUNNING', ~maStPtr, maCtx,$\mathtt{\dots}$)
, StateMa(~fwId, 'IDLE', ~vmpck, nonce, maId,dots)
, VMPCK(~fwId, ~maId, ~vmpck)dots  $]$
\end{tcbTamarin2}
\caption{Migration Agent initialization and launch}
\label{fig:fwlaunchesma}
\end{figure}




The rule produces the state facts that establish a secure communication channel
between \snpfw{} and \snpma{}. More precisely, the persistent fact
\tlst{!StateFwMa(dots)} represents the part of the \snpfw{} thread state relevant to launching guests
and enforcing migration policies. It contains the \snpma{} thread context \tlst{maCtx}, which is read
and updated during execution (e.g., to increment the message counter).
Both that fact and the linear
fact \tlst{StateMa(dots)} include a state name (e.g., \tlst{'IDLE'}) as a
parameter that is updated as the state
machine is executed (cf.~\Cref{fig:MASM}).
The rule also outputs a \tlst{VMPCK(dots)} fact, which enables the adversary
to corrupt and extract the communication key.

\subsection{Guest Launch}


A simplified guest launch procedure is depicted in~\Cref{fig:ASPFWLaunchSM}.
An SNP-protected guest (\snpgvm{}) is launched using the sequence of rules \tlst{FwProvisionGvm},
\tlst{FwAssociateGvmMa} and \tlst{FwLaunchGvmAssocMa}, or the sequence of rules
\tlst{FwProvisionGvm} and \tlst{FwLaunchGvmNoAssocMa}, depending on whether migration is allowed or not (\tlst{gvmMigPolicy}).
The launch procedure has similarities to that of \snpma{}; here we only emphasize the differences.


\snpfw{} reads an image (\tlst{$\dollar$image}), an ID Block (\tlst{idBlock}),
and an ID Authentication Information Structure (\tlst{idAuth}) from the guest
owner (\snpgo{}) using the \tlst{FwProvisionGvm} rule.
An \tlst{idBlock} includes the migration policy (\tlst{gvmMigPolicy}) for the guest.
An \tlst{idAuth} pair
consists of an ID block signature (\tlst{idBlockSig}) and the public part of a
\snpgo{}-provided identity key (\tlst{pubIdk}), which is used to produce the signature.
The \snpfw{} validates the launch digest (as described
in~\Cref{sec:measurement}) and ensures signature verification using the
\tlst{Eq(dots)} action fact  and \tlst{Equality}
restriction.

\begin{tcbTamarin2}
restriction Equality:
$\mkwd{\forall}$ x y #i. Eq(x, y)@i $\mkwd{\Rightarrow}$ x = y
\end{tcbTamarin2}

The \snpfw{} installs in a guest context \tlst{gvmCtx} several (fresh) values, including:
an Offline Encryption Key \tlst{oek}, which is used to encrypt
contents of guest pages that have been swapped-out;
a Virtual Machine Root Key \tlst{vmrk}, which the derived keys may be rooted in;
a pointer \tlst{gvmStPtr} to a guest state which is freshly generated for each guest state machine transition; a
platform-launch identifier \tlst{pid} which persists throughout the lifetime of a guest on
a particular platform;
and a session identifier \tlst{sid} which is freshly generated upon each guest swap.
Additionally, we use the variable \tlst{gvmIsMig} as a flag to indicate
whether the guest has been migrated or not, and the variable \tlst{authTag} to store the
authentication tag of the swapped-out guest state.


The specification mandates that the firmware rejects the pages donated by a
hypervisor via an \vlst{SNP_GCTX_CREATE} call if they are not in the
\vlst{Firmware} state. Given that a
single \snpma{} thread may manage multiple \snpgvm{}s concurrently, each
\snpma{} message includes a guest context address to distinguish individual
guests, as can be seen in the \Cref{fig:MASM}.
 We link a guest context with an address
$\tlst{$\dollar$gctxAddr}$ and ensure that it is not already assigned on the
platform (\tlst{chipId}) by applying the \tlst{Unique} restriction; otherwise,
the adversary can replay the \tlst{MSG_VMRK_REQ} guest message,
prompting \snpfw{} to install the same \tlst{vmrk} in multiple contexts.


The produced state facts \tlst{StateGvm(dots)} and \tlst{StateFwGvm(dots)}
allow \snpgvm{} to request remote attestation and key derivation services via a secure communication channel,
and \snpfw{} to provide these services (note that \snpfw{} uses two separate state facts for execution: \tlst{StateFwGvm(dots)} and
\tlst{!StateFwMa(dots)}).
The two states are bound by \snpfw{}, who simply installs the \tlst{reportId} of \snpma{}
into the guest context as \tlst{maReportId}. Whenever \snpfw{} receives a request from \snpma{} to manage a guest,
\snpfw{} checks whether the \tlst{reportId} in
the \snpma{} context matches the \tlst{maReportId} in the \snpgvm{} context and
refuses to service the request otherwise.
The check is performed by the rule \tlst{FwLaunchGvmAssocMa},
wherein \snpfw{} receives the \tlst{vmrk} request from the \snpma{} and finalizes the guest launch procedure.

\begin{figure*}[h]
  \centering
  \begin{tikzpicture}[shorten >=1pt,node distance=2cm,auto]
    \tikzset{every node/.style={font=\scriptsize{}}}

    \tikzstyle{ma state} = [
    draw,
    thick,
    row sep=1mm,
    text width = 3.5cm,
    rounded corners,
    ]

    \tikzstyle{ma state name} = [
    text centered,
    inner sep = 0pt,
    outer sep = 0pt,
    text=darkgray,
    ]
    \tikzstyle{ma state desc} = [
    inner sep = 1pt,
    ]

    \matrix [draw, rounded corners,thick, inner sep = 1pt, text width = 1cm, fill=gray, ] (ma) at (2,1)
    {
      \node[text centered, thick, text=white] {$\mathsf{MA}$}; \\
    };

    \matrix [ma state,] (idle) [below = 0.4cm of ma]
    {
      \node[ma state name] {\smsn{IDLE}}; \\
    };

    \matrix [ma state, text width = 3.1cm] (VMRK-request) [below = 0.4cm of idle]
    {
      \node[ma state name] {\smsn{RUNNING}};\\
      \node[ma state desc] {
          Generate \tlstt{vmrk}\\
          Write \msgcolor{\tlstt{MSG\_VMRK\_REQ}}\\
          Include \tlstt{gctxAddr}
      };\\
    };

    \matrix [ma state, text width = 3.5cm] (import-request) [right = 0.5cm of VMRK-request.east]
    {
      \node[ma state name] {\smsn{RUNNING}};\\
      \node[ma state desc] {
          Write \msgcolor{\tlstt{MSG\_IMPORT\_REQ}}\\
          Include donated $\tlstt{gctxAddr}$
      };\\
    };

    \matrix [ma state, text width = 3.5cm] (export-request) [left = 0.5cm of VMRK-request.west]
    {
      \node[ma state name,] {\smsn{RUNNING}}; \\
      \node[ma state desc] {
        Write \msgcolor{\tlstt{MSG\_EXPORT\_REQ}}\\
        Include \tlstt{gctxAddr}\\
      }; \\
    };

    \matrix [ma state,] (export-response) [below = 0.4cm of export-request]
    {
      \node[ma state name] {\smsn{EXPORT\_REQUEST}}; \\
      \node[ma state desc] {
        Migrate \tlstt{gvmCtx} to associated \tlstt{MA}
        via established secure channel\\
      }; \\
    };

    \matrix [ma state, text width = 3.1cm] (VMRK-response) [below = 0.4cm of VMRK-request]
    {
      \node[ma state name] {\smsn{VMRK\_REQUEST}}; \\
      \node[ma state desc] {};\\
    };

    \matrix [ma state,] (import-response) [below = 0.4cm of import-request]
    {
      \node[ma state name] {\smsn{IMPORT\_REQUEST}}; \\
      \node[ma state desc] {};\\
    };

    \draw [-Latex,] (ma.south) --  (idle.north);

    \draw [-Latex,] (idle.south) --  (VMRK-request.north);

    \draw [-Latex,] (idle.west) -| (export-request.north);
    \draw [-Latex,]
    (idle.east)
    -| node[below right, text width = 3.2cm] {
      Read \tlstt{gvmCtx} from \tlstt{MA}
      and \tlstt{gctxAddr} from \tlstt{HV}
    }
    (import-request.north);

    \draw [-Latex,] (export-request)
    -- node[left] {Read \msgcolor{\tlstt{MSG\_EXPORT\_RSP}}} (export-response);

    \draw [-Latex,] (VMRK-request)
    -- node[right, ] {Read \msgcolor{\tlstt{MSG\_VMRK\_RSP}}} (VMRK-response);

    \draw [-Latex,] (import-request)
    -- node[right,] {Read \msgcolor{\tlstt{MSG\_IMPORT\_RSP}}} (import-response);

    \path (idle.north west) -- (idle.north) coordinate[pos=0.5] (idle-mid-nw);
    \draw [-Latex,] (export-response.south)
    |- ([yshift=-.2cm,xshift=-1.3cm]export-request.south west |- export-response.south west)
    |- ([yshift=0.4cm]idle.north west -| idle-mid-nw)
    -- (idle-mid-nw)
    ;

    \path (idle.north east) -- (idle.north) coordinate[pos=0.5] (idle-mid-ne);
    \path (idle.north east) -- (idle-mid-ne) coordinate[pos=0.5] (idle-quater-ne);
    \draw [-Latex,] (import-response.south)
    |- ([yshift=-.2cm,xshift=1.3cm]import-request.south east |- import-response.south east)
    |- ([yshift=0.4cm]idle.north east -| idle-quater-ne)
    -- (idle-quater-ne)
    ;

    \draw [-Latex,] (VMRK-response.south)
    |- ([yshift=-.33cm,xshift=1.4cm]import-request.south east |- import-response.south east)
    |- ([yshift=0.53cm]idle.north east -| idle-mid-ne)
    -- (idle-mid-ne)
    ;

  \end{tikzpicture}
  \caption{Migration Agent State Machine}
  \label{fig:MASM}
\end{figure*}

\subsection{Guest messages}
\label{sec:guest_messages}

SNP-protected guests can utilize the \vlst{SNP_GUEST_REQUEST} command to securely
communicate with the AMD-SP firmware. All exchanged messages are tagged, encrypted and
integrity-protected using AES-256-GCM with the Virtual Machine Platform
Communication Key (\vlst{vmpck}) that is injected into guest pages through the
\vlst{SNP_LAUNCH_UPDATE} call. Guests, whether primary or migration agents,
employ messages for tasks like obtaining attestation reports, deriving keys,
and managing migration.

We use the built-in \tlst{natural-numbers} theory in \tamarin{} to represent nonces. The theory features the
constant \tlst{1:nat} and an associative-commutative union operator \tlst{\%+}, which suffice to model nonces via
monotonically increasing counters.

\snpgvm{} and \snpma{} each maintain their own message counter (\tlst{msgCount}), while
\snpfw{} maintains two separate message counters---one for \snpgvm{} and \snpma{} each.
They are initialized to \tlst{1:nat} and incremented
upon each successful message receipt (as in, e.g., \tlst{FwLaunchGvmAssocMa}).

\subsection{Guest Context}


\snpfw{} threads use \tlst{!StateFwMa(dots)} state facts to store contexts of \snpma{} threads for the purpose of guest management.  In addition, each \snpfw{}
thread may spawn any number of \tlst{StateFwGvm(dots)} state facts used to
store the contexts of \snpgvm{} threads, which enables the provision of services to guests.

Note that we represent each guest management state by a
persistent fact; otherwise, if we use a linear fact, then in a rule which involves reading the state, such as \tlst{FwProvisionGvm}, the same fact must appear on both sides of the rule (if the rule needed to update the state instead, the fact on the right-hand side would contain the updated values). This results in non-termination when inductively proving
some true statements, such as \tlst{SupNoKeyNonceReuseMA}, where \tamarin{} perpetually loops over the \tlst{FwProvisionGvm} rule in an unsuccessful effort to apply the induction hypothesis.

However, using a persistent fact makes it harder to update guest management states (although a new version of the fact can be added to the global state, the old version can never be removed). Updates are required, for example, to increment a message counter when \snpfw{} services the \tlst{MSG_VMRK_REQ}
request via the \tlst{FwLaunchGvmAssocMa} rule. We can fortunately employ a trick used for this purpose by Cremers et al.~\cite{wpa2}: we generate a fresh value that serves as a
pointer to the persistent fact, i.e.\ the memory represented by the fact, which can then be read via \tlst{Read(dots)} actions and released via
\tlst{Free(dots)} actions. We can then simply allocate new memory by generating a new persistent fact and a fresh pointer to it.

To ensure that memory is
always read before it is released and that it cannot be released more than
once, we use the respective two restrictions.

\begin{tcbTamarin2}
restriction FreedMemoryCannotBeRead:
$\mkwd{\forall}$ ptr #i #j. Read(ptr)@i $\mkwd{\land}$ Free(ptr)@j $\mkwd{\Rightarrow}$ #i < #j

restriction MemoryCanBeFreedOnlyOnce:
$\mkwd{\forall}$ ptr #i #j. Free(ptr)@i $\mkwd{\land}$ Free(ptr)@j $\mkwd{\Rightarrow}$ #i = #j
\end{tcbTamarin2}

\subsection{Remote Attestation}

Each \snpgvm{} may send any number of attestation report requests (\tlst{MSG\_REPORT\_REQ} messages)
to \snpfw{} via the \tlst{GVMRequestsReport} rule, regardless of whether the guest is migrated or not.
Each request
includes the \tlst{reportData} field---an arbitrary value provided by the
\snpgvm{} and uninterpreted by the \snpfw{}.  The \snpgo{} utilizes it to ensure
attestation report freshness.
This is shown in~\Cref{fig:GVMSM} and~\Cref{fig:GOSM}.

While the specification accommodates guests with
an option to select either \vlst{VCEK} or \vlst{VLEK} as a report signing key,
we only ever use \vlst{VCEK} for that purpose (i.e., \tlst{privVcek} in the
model).


\snpfw{} assembles an attestation report, as illustrated in~\Cref{fig:ASPFWCommSM}, by utilizing the
\tlst{FWGeneratesReport} rule.  The report comprises various information from
the guest context including a chip identifier (\tlst{chipId}), a migration
policy (\tlst{gvmMigPolicy}), and a report identifier (\tlst{reportId}).  An
attestation report is always signed with \tlst{privVcek} and forwarded by \snpgvm{} to the public
network, where it is available for inspection by \snpgo{}.




\begin{figure*}
  \centering
  \begin{tikzpicture}[shorten >=1pt,node distance=2cm,auto]
    \tikzset{every node/.style={font=\scriptsize{}}}

    \tikzstyle{fw state} = [
    draw,
    thick,
    row sep=1mm,
    text width = 3.5cm,
    rounded corners,
    ]

    \tikzstyle{fw state name} = [
    text centered,
    inner sep = 0pt,
    outer sep = 0pt,
    text=darkgray,
    ]
    \tikzstyle{fw state desc} = [
    inner sep = 1pt,
    ]

    \matrix [draw, rounded corners,thick, inner sep = 1pt, text width = 1cm, fill=gray, ] (fw) at (2,1)
    {
      \node[text centered, thick, text=white] {$\mathsf{FW}$}; \\
    };

    \matrix [fw state, text width = 4.5cm] (aspfw-ma-idle) [below = 0.4cm of fw]
    {
      \node[fw state name] {\smsn{IDLE}}; \\
    };

    \matrix [fw state, text width = 5.0cm] (aspfw-ma-create-gvm)
    [below left = 0.4cm and -1.2cm of aspfw-ma-idle]
    {
      \node[fw state name] {\smsn{RUNNING}}; \\
      \node[fw state desc] {
        Install \tlstt{vmpck}, \tlstt{vmrk}, \tlstt{oek}, \tlstt{reportId} to \tlstt{gvmCtx}\\
        Verify \tlstt{sigIdBlock}\\
        Include \tlstt{idBlock} and \tlstt{idAuth} to \tlstt{gvmCtx}
      };\\
    };

    \matrix [fw state, text width = 4.5cm] (aspfw-ma-import)
    [below right = 0.4cm and -1.2cm of aspfw-ma-idle]
    {
      \node[fw state name] {\smsn{RUNNING}}; \\
      \node[fw state desc] {
        Generate and install \tlstt{reportId} to \tlstt{gvmCtx}\\
        Install \tlstt{gvmCtx} to guest state\\
        Write \tlstt{MSG\_IMPORT\_RSP}\\
      };\\
    };

    \matrix [fw state, text width = 4.5cm] (aspfw-swap-in)
    [below = 0.4cm of aspfw-ma-import]
    {
      \node[fw state name] {\smsn{IMPORT}}; \\
      \node[fw state desc] {
          Verify \tlstt{authtag $\ \mathtt{\stackrel{?}{=}}\ $  gvmCtx.authTag}\\
          \tlstt{state = sdec(encSnapshot, oek)}\\
          Execute \tlstt{GVM}
      };\\
    };

    \matrix [fw state, text width = 4.5cm] (aspfw-launch-no-assoc)
    [below left = 0.4cm and -2.6cm of aspfw-ma-create-gvm]
    {
      \node[fw state name] {
        \smsn{INITIALIZE}
      }; \\
      \node[fw state desc] {
        Execute \tlstt{GVM}
      };\\
    };

    \matrix [fw state, text width = 4.5cm] (aspfw-ma-assoc)
    [below right = 0.4cm and -2.6cm of aspfw-ma-create-gvm]
    {
      \node[fw state name] {
        \smsn{INITIALIZE}
      }; \\
      \node[fw state desc] {
        Associate \tlstt{GVM} with \tlstt{MA}\\
        Install \tlstt{MA}'s \tlstt{reportId} to \tlstt{gvmCtx}\\
      };\\
    };

    \matrix [fw state, text width = 4.5cm] (aspfw-ma-install-VMRK)
    [below = 0.4cm of aspfw-ma-assoc]
    {
      \node[fw state name] {
        \smsn{ASSOCIATE}
      }; \\
      \node[fw state desc] {
        Verify \tlstt{reportId $\ \mathtt{\stackrel{?}{=}}\ $  gvmCtx.maReportId}\\
        Replace \tlstt{vmrk} in \tlstt{gvmCtx} with \tlstt{MA}'s\\
        Write \tlstt{MSG\_VMRK\_RSP}\\
        Execute \tlstt{GVM}
      };\\
    };

    \draw [-Latex,] (fw.south) --  (aspfw-ma-idle.north);

    \path (aspfw-ma-idle.south west) -- (aspfw-ma-idle.south) coordinate[pos=0.2] (aspfw-idle-sw);
    \draw [-Latex,]
    (aspfw-idle-sw) -- node[left] {Receive \tlstt{DEPLOY\_REQ}}
    (aspfw-idle-sw|- aspfw-ma-create-gvm.north);

    \path (aspfw-ma-idle.south east) -- (aspfw-ma-idle.south) coordinate[pos=0.2] (aspfw-idle-se);
    \draw [-Latex,]
    (aspfw-idle-se) -- node[right] {Read \tlstt{MSG\_IMPORT\_REQ}}
    (aspfw-idle-se|- aspfw-ma-import.north);

    \path (aspfw-ma-create-gvm.south west) -- (aspfw-ma-create-gvm.south) coordinate[pos=0.5] (aspfw-ma-create-gvm-sw);
    \draw [-Latex,]
    (aspfw-ma-create-gvm-sw) -- node[left] {\tlstt{gvmMigPolicy = 0}}
    (aspfw-ma-create-gvm-sw|- aspfw-launch-no-assoc.north);

    \path (aspfw-ma-create-gvm.south east) -- (aspfw-ma-create-gvm.south) coordinate[pos=0.5] (aspfw-ma-create-gvm-se);
    \draw [-Latex,]
    (aspfw-ma-create-gvm-se) --node[right] {\tlstt{gvmMigPolicy = 1}}
    (aspfw-ma-create-gvm-se|- aspfw-ma-assoc.north);

    \draw [-Latex,] (aspfw-ma-assoc.south) --
    node[right] {Read \tlstt{MSG\_VMRK\_REQ}}
    (aspfw-ma-install-VMRK.north);

    \draw [-Latex,] (aspfw-ma-import.south) --
    node[left] {Read \tlstt{SNP\_PAGE\_SWAP\_IN}}
    (aspfw-swap-in.north);

    \path (aspfw-ma-idle.north west) -- (aspfw-ma-idle.north) coordinate[pos=0.5] (idle-ma-mid-nw);
    \path (aspfw-ma-idle.north west) -- (idle-ma-mid-nw) coordinate[pos=0.5] (idle-ma-quarter-nw);

    \draw [-Latex,] (aspfw-launch-no-assoc.south)
    |- ([yshift=-.2cm,xshift=-0.2cm]aspfw-launch-no-assoc.south west |- aspfw-launch-no-assoc.south west)
    |- ([yshift=0.4cm]aspfw-ma-idle.north west -| idle-ma-quarter-nw)
    -- (idle-ma-quarter-nw)
    ;

    \draw [-Latex,] (aspfw-ma-install-VMRK.south)
    |- ([yshift=-.2cm,xshift=-0.4cm]aspfw-launch-no-assoc.south west |- aspfw-ma-install-VMRK.south west)
    |- ([yshift=0.6cm]aspfw-ma-idle.north west -| idle-ma-mid-nw)
    -- (idle-ma-mid-nw)
    ;

    \path (aspfw-ma-idle.north east) -- (aspfw-ma-idle.north) coordinate[pos=0.5] (idle-ma-mid-ne);
    \draw [-Latex,] (aspfw-swap-in.south)
    |- ([yshift=-.2cm,xshift=0.4cm]aspfw-swap-in.south east |- aspfw-swap-in.south east)
    |- ([yshift=0.4cm]aspfw-ma-idle.north west -| idle-ma-mid-ne)
    -- (idle-ma-mid-ne)
    ;

  \end{tikzpicture}
  \caption{AMD Security Processor Firmware State Machine (guest launch)}
  \label{fig:ASPFWLaunchSM}
\end{figure*}

\subsection{Key Derivation}

\snpgvm{} may also ask \snpfw{} to derive and provide a key by sending the
\tlst{MSG_KEY_REQ} guest message, as depicted in~\Cref{fig:GVMSM}, using the
\tlst{GVMRequestsKey} rule.  The keys
are derived by applying the ternary function symbol \tlst{kdf} to a root key and
additional data. This is also shown in~\Cref{fig:ASPFWCommSM}.

The root key is the \tlst{vmrk} of \snpma{} or \snpfw{},
depending on whether \snpgvm{} is associated with \snpma{} or not.



The acquired key is deleted by \snpgvm{} before the subsequent request. However,
the adversary can gain knowledge of the \tlst{oek}-encrypted key if it swaps out
\snpgvm{} while it is still in the \tlst{IDLE} state.
In fact, the confidentiality of a derived key depends on the confidentiality of
five different keys as we will see later.

%



\subsection{Page Swap}

The model incorporates a swapping mechanism where an adversary may use
\tlst{SNP_PAGE_SWAP_OUT} and \tlst{SNP_PAGE_SWAP_IN} commands to swap out and
swap in \snpgvm{} any number of times. The corresponding state machine is visualized
in~\Cref{fig:ASPFWCommSM}.
We do not explicitly
model memory pages of guests or VM Encryption Keys (\vlst{VEK}) used to
encrypt them.  Instead, we represent \snpgvm{} memory through \tlst{StateGvm(dots)} state facts,
and we model decryption and encryption with \vlst{VEK}
through the consumption and production of the \tlst{StateGvm(dots)} state facts.
\snpfw{} utilizes four rules \tlst{FwSwapOutGvm}(\tlst{BM}) and
\tlst{FwSwapInGvm}(\tlst{BM}) to swap out and swap in \snpgvm{}; employing two
almost identical pairs of rules for each purpose (\tlst{BM} stands for Before Migration) facilitates our modelling of the migration
procedure.

We abstract away from the details of stream cipher encryption with the \vlst{OEK}. Instead, we use the built-in theory
\tlst{symmetric-encryption} which defines two binary function symbols \tlst{senc}
and \tlst{sdec} related by the equation \tlst{sdec(senc(m,k),k)=m}.
Moreover, we introduce an irreducible binary function symbol
\tlst{mac} for the purpose of producing authentication tags.

During the swap-out process, \snpfw{} encrypts the contents of the \snpgvm{} state with
\tlst{oek} to obtain the encrypted snapshot (\tlst{gvmEncSnapshot}) of guest's memory, which it subsequently uses to
calculate the authentication tag \tlst{authTag}. While both \tlst{gvmEncSnapshot} and
\tlst{authTag}  are then transmitted over the network, \tlst{authTag} is
saved in the \snpgvm{} context.
Additionally, \tlst{StateGvm(dots)} and \tlst{StateFwGvm(dots)} state facts are parameterized with a fresh session identifier \tlst{\~sid} to
facilitate backward search.

Conversely, during the swap-in process, \snpfw{} first receives \tlst{gvmEncSnapshot} and
\tlst{authTag} from the network. It then decrypts the \snpgvm{} state contents with
\tlst{oek} and verifies \tlst{authTag} by comparing it against the saved one
using the \tlst{Equality} restriction. If the check succeeds,
the \snpgvm{} thread may continue its execution afterwards.
We can demonstrate (as described in~\Cref{sec:executability}) that ignoring
this check leads to a rollback attack where the same key and nonce get reused.

For termination reasons, we prohibit the \snpfw{} from swapping out the same
\snpgvm{} state more than once.
Each \snpgvm{} transition generates a
fresh pointer, denoted as \tlst{gvmStPtr}, which uniquely identifies a particular
state. To ensure that the state is swapped out at most once, we enforce the following
restriction:

\begin{tcbTamarin2}
restriction MemoryCanBeSwappedOnlyOnce:
  $\mkwd{\forall}$ k1 k2 guest1 guest2 state1 state2 pointer #i #j.
    Swap(k1, guest1, state1, pointer)@i $\mkwd{\land}$
    Swap(k2, guest2, state2, pointer)@j $\mkwd{\Rightarrow}$ #i = #j
\end{tcbTamarin2}

We do not consider this to be a limitation since, in our model, the adversary
gains no advantage by obtaining the same ciphertext multiple times.

For performance reasons, we also prohibit swapping out \snpgvm{} while it is
in the \tlst{RUNNING} state by employing the following restriction:

\begin{tcbTamarin2}
restriction GuestStateDuringASwap:
$\mkwd{\forall}$ kind guest state pointer #i.
  Swap(kind, guest, state, pointer)@i $\mkwd{\Rightarrow}$ state = 'IDLE'
         $\mkwd{\lor}$ state = 'KEY_REQ' $\mkwd{\lor}$ state = 'REPORT_REQ'
\end{tcbTamarin2}

We also do not regard this as a limitation since a \snpgvm{} \tlst{IDLE}
state, which may be swapped out, differs from a \tlst{RUNNING}
state only in the deletion of a derived key (modelled by replacing the key with the \tlst{kdf('NULL', 'NULL', 'NULL')} fact).

\begin{figure*}
  \centering
  \begin{tikzpicture}[shorten >=1pt,node distance=2cm,auto]
    \tikzset{every node/.style={font=\scriptsize{}}}

    \tikzstyle{fw state} = [
    draw,
    thick,
    row sep=1mm,
    text width = 3.5cm,
    rounded corners,
    ]

    \tikzstyle{fw state name} = [
    text centered,
    inner sep = 0pt,
    outer sep = 0pt,
    text=darkgray,
    ]
    \tikzstyle{fw state desc} = [
    inner sep = 1pt,
    ]

    \matrix [draw, rounded corners,thick, inner sep = 1pt, text width = 1cm, fill=gray, ] (fw) at (2,1)
    {
      \node[text centered, thick, text=white] {$\mathsf{FW}$}; \\
    };

    \matrix [fw state,] (aspfw-gvm-idle) [below = 0.4cm of fw]
    {
      \node[fw state name] {\smsn{IDLE}}; \\
    };

    \matrix [fw state, text width = 4.3cm] (aspfw-swap-out) [below = 0.4cm of aspfw-gvm-idle]
    {
      \node[fw state name] {\smsn{RUNNING}};\\
      \node[fw state desc] {
          \tlstt{encSnapshot = senc(state, oek)}\\
          \tlstt{authTag = mac(encSnapshot, oek)}\\
          Save \msgcolor{\tlstt{authtag}} to \tlstt{gvmCtx}\\
          Write \msgcolor{\tlstt{encSnapshot}} and \tlstt{authTag}\\
      };\\
    };

    \matrix [fw state, text width = 3.8cm] (aspfw-report-request) [right = 0.5cm of aspfw-swap-out.east]
    {
      \node[fw state name] {\smsn{RUNNING}};\\
      \node[fw state desc] {
        Create \tlstt{report} from \tlstt{gvmCtx}\\
        \tlstt{sig = sign(report, VCEK)}\\
        Write \msgcolor{\tlstt{MSG\_REPORT\_RSP}}\\
      };\\
    };

    \matrix [fw state, text width = 4.6cm] (aspfw-key-request) [left = 0.5cm of aspfw-swap-out.west]
    {
      \node[fw state name,] {\smsn{RUNNING}}; \\
      \node[fw state desc] {
          \tlstt{root = vmrk}\\
          \tlstt{params = <vmpl, hostData, pubIdk>}\\
          \tlstt{key = kdf('VMRK', root, params)}\\
          Write \msgcolor{\tlstt{MSG\_KEY\_RSP}}\\
      }; \\
    };

    \matrix [fw state, text width = 4.4cm] (aspfw-swap-in)
    [below left = 0.4cm and -2.0cm of aspfw-swap-out]
    {
      \node[fw state name] {\smsn{SWAP\_OUT}}; \\
      \node[fw state desc] {
          Verify \tlstt{authtag $\ \mathtt{\stackrel{?}{=}}\ $  gvmCtx.authTag}\\
          \tlstt{state = sdec(encSnapshot, oek)}\\
          Execute \tlstt{GVM}
      };\\
    };

    \matrix [fw state, text width = 4.5cm] (aspfw-export-request)
    [below right = 0.4cm and -2.0cm of aspfw-swap-out]
    {
      \node[fw state name] {\smsn{SWAP\_OUT}}; \\
      \node[fw state desc] {
          Verify \tlstt{reportId $\ \mathtt{\stackrel{?}{=}}\ $  gvmCtx.maReportId}\\
          \tlstt{encGctx = wrap(gvmCtx, nonce, vmpck)}
          Write \tlstt{MSG\_EXPORT\_RSP}\\
      };\\
    };

    \draw [-Latex,] (fw.south) --  (aspfw-gvm-idle.north);

    \draw [-Latex,] (aspfw-gvm-idle.south) --
    node[right] {Read \msgcolor{\tlstt{SNP\_PAGE\_SWAP\_OUT}}}
    (aspfw-swap-out.north);

    \draw [-Latex,]
    (aspfw-gvm-idle.west)
    -| node[below left,] {
      Read \msgcolor{\tlstt{MSG\_KEY\_REQ}}
    }
    (aspfw-key-request.north);

    \draw [-Latex,]
    (aspfw-gvm-idle.east)
    -| node[below right] {
      Read \msgcolor{\tlstt{MSG\_REPORT\_REQ}}
    }
    (aspfw-report-request.north);

    \path (aspfw-gvm-idle.north west) -- (aspfw-gvm-idle.north) coordinate[pos=0.5] (idle-mid-nw);
    \path (aspfw-gvm-idle.north west) -- (idle-mid-nw) coordinate[pos=0.5] (idle-quarter-nw);

    \path (aspfw-swap-out.south west) -- (aspfw-swap-out.south) coordinate[pos=0.5] (aspfw-swap-out-mid-sw);
    \draw [-Latex,]
    (aspfw-swap-out-mid-sw) --
    node[left, ] {Read \msgcolor{\tlstt{SNP\_PAGE\_SWAP\_IN}}}
    (aspfw-swap-out-mid-sw |- aspfw-swap-in.north);

    \path (aspfw-swap-out.south east) -- (aspfw-swap-out.south) coordinate[pos=0.5] (aspfw-swap-out-mid-se);
    \draw [-Latex,]
    (aspfw-swap-out-mid-se) --
    node[right, ] {Read \msgcolor{\tlstt{MSG\_EXPORT\_REQ}}}
    (aspfw-swap-out-mid-se |- aspfw-export-request.north);

    \draw [-Latex,] (aspfw-key-request.south)
    |- ([yshift=-.2cm,xshift=-0.2cm]aspfw-key-request.south west |- aspfw-key-request.south west)
    |- ([yshift=0.4cm]aspfw-gvm-idle.north west -| idle-quarter-nw)
    -- (idle-quarter-nw)
    ;

    \draw [-Latex,] (aspfw-swap-in.south)
    |- ([yshift=-.2cm,xshift=-0.3cm]aspfw-key-request.south west |- aspfw-swap-in.south west)
    |- ([yshift=0.53cm]aspfw-gvm-idle.north west -| idle-mid-nw)
    -- (idle-mid-nw)
    ;

    \path (aspfw-gvm-idle.north east) -- (aspfw-gvm-idle.north) coordinate[pos=0.5] (idle-mid-ne);
    \draw [-Latex,] (aspfw-report-request.south)
    |- ([yshift=-.2cm,xshift=1.3cm]aspfw-report-request.south east |- aspfw-report-request.south east)
    |- ([yshift=0.4cm]aspfw-gvm-idle.north west -| idle-mid-ne)
    -- (idle-mid-ne)
    ;

  \end{tikzpicture}
  \caption{AMD Security Processor Firmware State Machine (guest management)}
  \label{fig:ASPFWCommSM}
\end{figure*}

\subsection{Live Migration}

SEV-SNP
offers several methods for migrating SNP-protected guests,
depending on whether the assistance of a migration agent or a guest is
utilized. In our model, we permit at most two migrations and do not consider guest-assisted migration.

The \snpma{} state machine is depicted in~\Cref{fig:MASM}.
The \snpma{} thread is tasked with providing a \tlst{vmrk} during the launch
and migrating guests to other platforms. Each migratable \snpgvm{},
as indicated by the \tlst{gvmMigPolicy} flag, is assigned to a single \snpma{}
on a particular platform. Conversely, a single \snpma{} thread
can manage an arbitrary number of primary guests. \snpma{} itself is not
migratable.

As mandated by the specification, \snpgvm{} is swapped out prior to migration
using the \tlst{FwSwapOutGvmBM} rule. Migration is then initiated by
\snpma{}, which sends the \tlst{MSG_EXPORT_REQ} guest message to \snpfw{}. Upon
receipt of the message via the \tlst{FwExportGvm} rule, the \snpfw{} verifies whether the
\snpma{} thread is assigned to the specific guest by comparing \tlst{reportId} in the
context of the \snpma{} thread with \tlst{maReportId} in the context of \snpgvm{}. Assuming
they are equal, \snpfw{} responds by transmitting the guest context via
\tlst{MSG_EXPORT_RSP} guest message. The exported context encompasses all
\snpgvm{} data except for \tlst{maReportId}, which will be replaced with the
\tlst{reportId} value of \snpma{} on the destination machine.

The specification indicates that the context is sent to a migration agent on the
destination machine through a secure channel. However, unlike in previous SEV
instances, the specific mechanism by which this transmission is achieved is outside the scope of the specification.

In our model, we assume that each \snpma{} thread on the source machine (source
\snpma{}), denoted by \tlst{maId}, is capable of establishing a secure
communication channel with an \snpma{} thread on the destination machine (target
\snpma{}). This communication channel is
modeled via two rules, \tlst{ComChannelOut} and \tlst{ComChannelIn}.

Upon receiving the \snpgvm{} context from the source \snpma{}, the target
\snpma{} initiates the import procedure by sending the \tlst{MSG_IMPORT_REQ}
guest message to the target \snpfw{}. Subsequently, the \snpfw{} utilizes the
\tlst{FwImportGvm} rule to add the guest context to the
\tlst{StateFwGvm(dots)} state fact and update it with a freshly generated
\tlst{reportId}. Additionally, the \snpfw{} establishes the association between
\snpma{} and \snpgvm{} by incorporating the \tlst{reportId} value of the \snpma{} into the
guest context as \tlst{maReportId}. Following this, \snpgvm{} is swapped in
using the \tlst{FwSwapInGvmAM} rule and launched afterward.

\newcommand{\nparbox}[2]{\parbox{#1}{\scriptsize{}#2}}
\newcommand{\msctext}[1]{\scriptsize{}#1}

\subsection{Secure Channel}

Migration agents employ \tlst{ComChannelOut} and \tlst{ComChannelIn} rules,
illustrated in~\Cref{fig:comchannel}, to
enable the secure transfer of guest contexts between platforms. This process
involves using \tlst{ComChan__(dots)} state facts to link 
\tlst{OutChan(dots)} and \tlst{InChan(dots)} state facts, for the purpose of
sending and receiving a guest context.  This use of state facts ensures that the
adversary can neither modify nor learn messages that are sent over the channel.
Furthermore, the use of linear facts ensures that the messages sent cannot be
replayed at a later point in time.

\begin{figure}
\begin{multicols}{2}

\begin{tcbTamarin2}
rule ComChannelOut:
$[$ OutChan(~A, ~B, m) $]$
$\mathtt{\mathrel{-}\xjoinrel\mathrel{[}}$ ComChanOut(~A, ~B, m) $\mathtt{\mathrel{]}\xjoinrel\rightarrow}$
$[$ ComChan__(~A, ~B, m) $]$
\end{tcbTamarin2}

  \columnbreak
\begin{tcbTamarin2}
rule ComChannelIn:
$[$ ComChan__(~A, ~B, m) $]$
$\mathtt{\mathrel{-}\xjoinrel\mathrel{[}}$ ComChanIn(~A, ~B, m) $\mathtt{\mathrel{]}\xjoinrel\rightarrow}$
$[$ InChan(~A, ~B, m) $]$
\end{tcbTamarin2}

\end{multicols}
\caption{Secure communication channel establishment}
\label{fig:comchannel}
\end{figure}



The \tlst{-DENABLE_REPLAY_OVER_COMM} flag selects a model variant
that permits message replay over the channel. In this model, a persistent fact
\tlst{!ComChan__(dots)} is used instead of a linear fact \tlst{ComChan__(dots)}.
We demonstrate how the adversary may exploit this to its advantage and
compromise \tlst{vmpck} by utilizing the
\tlst{ExeAttackReplayOverCommChan} executable lemma.

\subsection{Adversarial Model}

In accordance with the AMD SEV-SNP threat model, we assume that the adversary
has control over a hypervisor and a cloud provider and is able to launch an arbitrary
number of SNP-protected VMs, issue ABI commands in any order, spy on the
communication outputs, and tamper with the communication inputs.  Moreover, the
adversary may corrupt both the SNP firmware and guests, and extract keys
from repeated keystreams. The Dolev-Yao adversary of \tamarin{} has most of these
functionalities built in by default; we only had to manually enable the last two behaviors.

We introduce various \tlst{Extract} (such as the one in~\Cref{fig:extractVMPCK}) and \tlst{Reveal} rules that disclose, respectively,
secrets generated during launch and long-term keys to the adversary.
For instance, the adversary may use the \tlst{ExtractVmpck} rule
to corrupt either \snpgvm{} or \snpma{} and
extract \tlst{vmpck}.



\begin{figure}
\begin{tcbTamarin2}
rule ExtractVmpck:
 $[$ VMPCK(fwId, gvmId, key) $]$
$\mathrel{-}\xjoinrel\mathrel{[}$ CorruptVmpck(key),dots $\mathrel{]}\xjoinrel\rightarrow$
 $[$ Out(key) $]$
\end{tcbTamarin2}
\caption{VM Platform Communication Key compromise}
\label{fig:extractVMPCK}
\end{figure}

We allow the compromise of any key generated during launch,
excluding a specific finite set of keys bound to a particular thread
identifier (e.g., \tlst{gvmId}). Due to the swapping and migration
mechanisms, the confidentiality of a given key may rely on that of other keys, thereby introducing additional potential attack vectors.

Similarly, the adversary may employ the \tlst{ExtractCek} rule to extract
\tlst{cek} from \snpfw{}, which it might be able to do in practice by using side-channel attacks. It can also reveal the AMD Root Key \tlst{privArk} and the AMD Signing Key \tlst{privAsk}.



The Galois/Counter mode stream cipher is extensively utilized within AMD SEV-SNP.
However, considering the fact that the guest state may be swapped and
the guest context migrated, it is not clear whether this cipher is always correctly employed. We allow the adversary to recover the plaintext from one of two distinct guest messages, both encrypted with the same key and the same nonce.


\begin{figure}
\begin{tcbTamarin2}
rule MsgRevealFromKeyNonceReuse:
let encM1 = wrap(m1, nonce, key)
     encM2 = wrap(m2, nonce, key)
in
$[$ In(< encM1, encM2 >) $]$
$\mathrel{-}\xjoinrel\mathrel{[}$ Neq(m1, m2)
 , ReuseNonceKey(nonce, key) $\mathrel{]}\xjoinrel\rightarrow$
$[$ Out(m1) $]$
\end{tcbTamarin2}
\caption{Extracting messages from repeated keystreams}
\label{fig:keyextractfromnoncereuse}
\end{figure}


Note that we exclusively model the stream cipher for guest
messages, even though it is also employed for guest pages in SEV-SNP.



\subsection{Summary}

Our model includes 43 rewrite rules specified in
1,100 lines of code (LoC) without comments.
It has several constraints:

\begin{itemize}
  \item two migrations per guest are permitted;
  \item the stream cipher is only employed for guest messages;
  \item firmware updates are not supported;
  \item key derivation always utilizes the VM Root Key;
  \item we do not model the Versioned Loaded Endorsement Key.
\end{itemize}


\section{Analysis}\label{sec:analysis}

We classify the specified properties, also referred to as lemmas, into seven
distinct groups: \emph{source}, \emph{executability}, \emph{supporting},
\emph{secrecy}, \emph{authentication}, \emph{attestation}, and \emph{freshness}
lemmas. The corresponding counts of lemmas and the analysis time for each
group are outlined in Table~\ref{table:propsummary}.

\subsection{Executability Lemmas}
\label{sec:executability}

Executability lemmas claim the existence of traces of a particular form and are identified with the ``\tlst{exists-trace}''
keyword. The use of this keyword instructs \tamarin{} to mark the lemma as true if
there exists at least one execution which satisfies the underlying formula. This search for a particular trace
is in contrast with the security lemmas, which must hold for all possible executions.

We use executability lemmas in two ways. First, we establish the functional correctness of our model by verifying them. Namely, these lemmas encompass a range of potential behaviors within the model.
For instance, the \tlst{ExeMAManagesTwoGVMs} lemma specifies that a single
\snpma{} thread can be associated with two \snpgvm{}s.
In each such lemma we limit the number of rule instances and
prohibit any key compromise to reduce search time.

Second, we prove the existence of attacks in seemingly vulnerable model variants.
The rationale behind this is to verify whether certain checks and assumptions
are indeed necessary for security properties to hold.
Specifically, we employ the \vlst{m4} flags
\tlst{-DIGNORE_ROOT_MD_ENTRY} and \tlst{-DENABLE_REPLAY_OVER_COMM}, respectively, to disable
\tlst{authTag} verification during swap in and to enable replaying messages via
\tlst{CommChan} rules. Both lead to attacks
where the same key and nonce get reused.
We leverage the \tlst{ExeAttack} lemmas, such as
\tlst{ExeAttackSwapInRollbackBM}, to demonstrate this.

\subsection{Supporting Lemmas}


Due to the complexity of our model, including loops which tend to make unbounded
verification challenging, proving most of the security properties directly is
not feasible.
Therefore, we employ supporting lemmas to

\begin{itemize}
  \item provide entry points for loops;
  \item enforce termination;
  \item improve verification time.
\end{itemize}


We allow loops to execute an unbounded number of times; this models for instance
the perpetual servicing of \snpgvm{} requests by \snpfw{} and may lead to
non-termination of backward search. We can remedy this
by proving that each loop
has an entry point,
and we do so
through inductive reasoning.
See for instance the supporting lemma \tlst{SupFWInitializesGCTX}.


We also leverage supporting lemmas to prove other supporting and secrecy lemmas.
For instance, we may prove certain invariants that hold for every loop
iteration. An example of such a property is the lemma
\tlst{SupGVMMessageCounterMustBeEven}, which states that the \snpgvm{} message
counter always has an even value.
We utilize proof oracles to guide the \tamarin{} proof
procedure in a direction where such lemmas can be applied.


Most of the supporting lemmas are used to prove the
following lemma which states that
honest agents will never reuse a \tlst{nonce} with the same encryption \tlst{key} (i.e., \tlst{vmpck}):
\begin{tcbTamarin2}
lemma SupNoMsgRevealFromNonceReuse:
$\mkwd{\forall}$ nonce key #i. ReuseNonceKey(nonce, key)@i
                   $\mkwd{\Rightarrow}$ $\mkwd{\exists}$ #j. KU(key)@j $\mkwd{\land}$ #j < #i
\end{tcbTamarin2}

The difficulty arises as \tamarin{} considers all possible traces wherein two distinct
messages are encrypted with the same \tlst{vmpck} using the same \tlst{nonce}.
Moreover, the ability to swap a guest any number of times further complicates
matters. The \tlst{SupNoMsgRevealFromNonceReuse} lemma is necessary to prove most of the secrecy lemmas.

The fresh variable \tlst{pid} is used to enforce that
\tlst{StateFwGvm} and \tlst{StateGvm} fact symbols have \emph{injective
instances}~\cite{tamarinmanual}.  Our analysis relies on the capability of
\tamarin{} to underapproximate a set of such symbols. For instance, the
following lemma can be proven only if we assume that
\tlst{StateGvm} has injective instances:

\begin{tcbTamarin2}
lemma SupNoGVMHandlingAfterSwapOut:
$\mkwd{\lnot}$ ($\mkwd{\exists}$ pid #i #j #k dots.
       FwActivateGuest(pid)@i
    $\mkwd{\land}$ FwDeactivateGuest(pid, dots)@j
    $\mkwd{\land}$ FwHandleGvm(dots, pid, dots)@k
    $\mkwd{\land}$ (#i < #j)
    $\mkwd{\land}$ (#j < #k))
 \end{tcbTamarin2}

This lemma states that no guest request handling is possible after the guest has
been swapped out, prior to export.  Most of the subsequent supporting lemmas,
including the previously mentioned \tlst{SupNoMsgRevealFromNonceReuse}, depend
on this property.

\subsection{Secrecy Lemmas}

\begin{lemma*}
  \centering
  \begin{sublemma}{.48\textwidth}
    \begin{tcbTamarin2}
lemma SecKeyDerFromMaVmrkIsSecret:
$\mkwd{\forall}$ gctxAddr gvmId key info params fwId vmpck maId1 maId2
   vmrk #i #j #k #l.
   AssociateMigrationAgent(maId1, maId2)@i
$\mkwd{\land}$  MaGenerateVmrk(maId, fwId, gctxAddr, vmrk)@j
$\mkwd{\land}$  FwInstallMaVmrk(fwId, vmpck, maId, gvmId,
                      gctxAddr, vmrk)@k
$\mkwd{\land}$ key = kdf(info, vmrk, params)
$\mkwd{\land}$ KU(key)@l
$\mkwd{\Rightarrow}$    ($\mkwd{\exists}$ #m. (#m < #l) $\mkwd{\land}$ CorruptVmrk(vmrk)@m)
 $\mkwd{\lor}$   ($\mkwd{\exists}$ #m. (#m < #l) $\mkwd{\land}$ CorruptVmpck(vmpck)@m)
 $\mkwd{\lor}$   ($\mkwd{\exists}$ #m. (#m < #l) $\mkwd{\land}$ CorruptImageOek(gvmId)@m)
 $\mkwd{\lor}$   ($\mkwd{\exists}$ #m. (#m < #l) $\mkwd{\land}$ CorruptImageVmpck(maId1)@m)
 $\mkwd{\lor}$   ($\mkwd{\exists}$ #m. (#m < #l) $\mkwd{\land}$ CorruptImageVmpck(maId2)@m)
    \end{tcbTamarin2}
    \lemmanewsubcap{Secrecy of a \tlst{vmrk}-derived key}
    \label{lem:sec}
  \end{sublemma}%
\hspace*{10pt}
  \begin{sublemma}{.48\textwidth}
    \begin{tcbTamarin2}
lemma AuthFwGvmAgreeMsgAttest:
$\mkwd{\forall}$ isMig gvmId chipId fwId vmpck
   maId1 maId2 msg #i #j #k.
   AssociateMigrationAgent(maId1, maId2)@i
$\mkwd{\land}$ FwAssociateMaGvm(gvmId, maId1)@j
$\mkwd{\land}$ FwReceiveGvmRequest('FW_GENERATE_REPORT', isMig,
            chipId, fwId, gvmId, vmpck, msg)@k
$\mkwd{\Rightarrow}$ ( $\mkwd{\exists}$ #l. (#l < #k)
      $\mkwd{\land}$ GvmIssueRequest('GVM_REPORT_REQUEST', isMig,
            gvmId, chipId, fwId, msg)@l )
 $\mkwd{\lor}$ ($\mkwd{\exists}$ #l. (#l < #k) $\mkwd{\land}$ CorruptVmpck(vmpck)@l)
 $\mkwd{\lor}$ ($\mkwd{\exists}$ #l. (#l < #k) $\mkwd{\land}$ CorruptImageOek(gvmId)@l)
 $\mkwd{\lor}$ ($\mkwd{\exists}$ #l. (#l < #k) $\mkwd{\land}$ CorruptImageVmpck(maId1)@l)
 $\mkwd{\lor}$ ($\mkwd{\exists}$ #l. (#l < #k) $\mkwd{\land}$ CorruptImageVmpck(maId2)@l)
    \end{tcbTamarin2}
    \lemmanewsubcap{Agreement on the \tlst{MSG_REPORT_REQ} guest message}
    \label{lem:auth}
  \end{sublemma}
\end{lemma*}

We verify the perfect forward secrecy of each key that we use within the
model, including long-term, generated, and derived keys.
In all security properties, we permit the adversary to corrupt dishonest
agents, and to take advantage of key and nonce reuse of any agent.

We specify security properties depending on whether the guest is migratable or not. This approach enables us to verify certain properties in the presence of a more
powerful adversary.
For instance, when analyzing the secrecy of \tlst{oek}, we consider two lemmas:
\tlst{SecGvmOekIsSecret} and \tlst{SecGvmOekIsSecretMd}. These lemmas
correspond to scenarios where the association with an \snpma{} is allowed or disallowed, respectively. The lemmas differ in that the latter allows the adversary to corrupt
even the \snpma{} thread that is running in the background on the same platform.

Key secrecy is usually contingent on several other keys. Take, for
instance, Lemma~\ref{lem:sec}, which specifies the secrecy of the \tlst{key} derived from
the \tlst{vmrk} of \snpma{}.
With this property, we consider the scenario wherein the \tlst{vmrk} is
generated by the \snpma{} thread \tlst{maId} and installed by the \snpfw{}
thread \tlst{fwId} within the context of the \snpgvm{} thread
\tlst{gvmId}.  If we assume that the adversary knows the key derived from
\tlst{vmrk}, then it must necessarily be the case that either
\tlst{vmrk}, \tlst{vmpck} or \tlst{oek} of \tlst{gvmId} is corrupted, or one of the
associated \snpma{}s is corrupted.

Here, the secrecy of the \tlst{key} critically depends on the secrecy of five other keys. To illustrate
this, consider the various way in which \tlst{key} could potentially be
compromised.

First, if the adversary corrupts the \tlst{vmrk}, it can clearly construct the
\tlst{key} independently. Second, the adversary can obtain the \tlst{key} by compromising the \tlst{vmpck} of the \snpgvm{} thread and decrypting the guest message \tlst{MSG_KEY_RSP}. Third, the adversary can swap
out the state of the \snpgvm{} thread and acquire either the \tlst{vmpck}
or \tlst{key} directly (assuming it has not been deleted and the adversary possesses the corresponding \tlst{oek}). Finally, corrupting the \snpma{} thread on either
the source or destination platform lets the adversary obtain \tlst{vmrk} from
a \tlst{MSG_EXPORT_REQ} or \tlst{MSG_IMPORT_REQ} guest message, respectively.


Note that most of the specified secrecy properties can additionally be viewed as supporting lemmas, because they facilitate the verification of authentication, attestation, freshness, and other secrecy lemmas.

\subsection{Authentication Lemmas}

We utilize authentication lemmas to verify whether the firmware and guest,
with the latter being either \snpgvm{} or \snpma{}, agree on the messages exchanged. Specifically, we
are interested in whether the messages received by the guest, running on a
particular platform, indeed originate from the firmware on that same platform,
and vice-versa.

Take for instance Lemma~\ref{lem:auth}, which specifies agreement on the guest
message \tlst{MSG_REPORT_REQ}. It considers a \snpgvm{} thread with the thread identifier
\tlst{gvmId}, launched by a \snpfw{} thread with the thread identifier
\tlst{fwId}, under a policy that permits migration.  Here, we want to verify
whether the attestation report \tlst{request} the \snpfw{} obtains was indeed issued by
the \snpgvm{} after launch, assuming both honest agents
are uncompromised.



\subsection{Attestation Lemmas}

The attestation lemmas specify authenticity and integrity properties of attestation reports.
Both kinds of property consider the scenario where \snpgo{} verifies an attestation report
using the signing key that apparently belongs to a particular chip.
The authenticity properties then affirm the existence of a \snpfw{}
thread, operating on that chip, which previously generated the same report.
In contrast, the integrity properties are concerned
with the state of the guest bound to that report,
asserting that the guest is indeed running on the designated platform, with
the correct policy and configuration.
Lemma~\ref{lem:attest} is an example of such a property.

\subsection{Freshness Lemmas}

The freshness lemmas comprise several \emph{uniqueness} lemmas.
In particular, these lemmas allow us to determine
whether two separate \snpgvm{} threads can acquire the same derived key, whether
\snpgo{} can verify outdated certificates, or if the launch procedure can be
manipulated to enforce \snpfw{} to install the same \tlst{vmrk} into multiple
guests.

As we mentioned previously, guests have the option, via \vlst{MSG_KEY_REQ}, to choose the root
key---\vlst{VCEK}, \vlst{VLEK}, or \vlst{vmrk}---and provide additional data for
key derivation. Opting for \vlst{vmrk} as the root key should ensure that each guest
instance will obtain a different key; see lemma
\tlst{FreshKeyDerFromFwVmrkIsGvmUnique}. We note here, however, that it is not possible for
a guest to provide a random sequence of bytes to be included in SEV-SNP key derivation (so unlike in Intel SGX, key wear-out protection is not supported).



\section{Results and Discussion}\label{sec:results}






We analyzed a total of 114 properties, including 42 security
properties. The summary of the results is
presented in Table~\ref{table:propsummary}. All but five security properties were
successfully verified.

The analysis of the model was conducted using Debian 11, running
on an AMD EPYC 7713 processor, equipped with 16 cores
and 32 threads, with a 2.0GHz base clock. The system was complemented by
256GB of memory. We employed \tamarin{} version 1.10.0 (commit hash
\vlst{cb62c305}) for the analysis.

All of the specified properties were automatically analyzed in
about 6 hours, as detailed in Table~\ref{table:propsummary}.
In this section, we discuss the results in detail, including the potential impact and mitigations for the discovered weaknesses.
These findings were disclosed to AMD.

\begin{lemma}
\begin{tcbTamarin2}
lemma AttReportStrongIntegrity:
$\mkwd{\forall}$ goId goImage privIdk report image policy reportData
  ld digestIdk pubIdk reportId maReportId chipId fwId
  hostData maId1 maId2 vmpck imageId gvmId fwLaunchTcb
  fwCurrentTcb askId arkId ltks #i #j #k #l.
   AssociateMigrationAgent(maId1, maId2)@i
$\mkwd{\land}$ FwInstallVmpck('GVM', fwId, gvmId, vmpck)@j
$\mkwd{\land}$ FwAssociateMaGvm(gvmId, maId1)@o
$\mkwd{\land}$ FwBindReportIdGvm(vmpck, reportId)@p
$\mkwd{\land}$ GoReportVerify(goImage, goId, privIdk, report, ltks)@l
$\mkwd{\land}$ ltks = <arkId, askId>
$\mkwd{\land}$ report = <imageId, policy, reportData, ld,
     digestIdk, reportId, maReportId, chipId, gvmId,
     fwLaunchTcb, fwCurrentTcb>
$\mkwd{\land}$ policy = 'ENABLE_MIGRATION'
$\mkwd{\land}$ digestIdk = h(<'ID_KEY', pubIDK>)
$\mkwd{\land}$ pubIdk        = pk(privIdk)
$\mkwd{\land}$ ld            = h(< 'VM_IMAGE', image >)
$\mkwd{\Rightarrow}$ $\mkwd{\exists}$ isMig vmpck gvmId gvmMsgCount #m.
      (#m < #l)
   $\mkwd{\land}$ GvmReportReq(chipId, fwId, imageId, gvmId, vmpck,
                     gvmMsgCount:nat, reportData, isMig)@m
$\mkwd{\lor}$ ($\mkwd{\exists}$ #m. (#m < #l) $\mkwd{\land}$ CorruptLtk('ARK', arkId)@m )
$\mkwd{\lor}$ ($\mkwd{\exists}$ #m. (#m < #l) $\mkwd{\land}$ CorruptLtk('ASK', askId)@m )
$\mkwd{\lor}$ ($\mkwd{\exists}$ #m. (#m < #l) $\mkwd{\land}$ CorruptLtk('CEK', chipId)@m )
$\mkwd{\lor}$ ($\mkwd{\exists}$ #m. (#m < #l) $\mkwd{\land}$ CorruptVmpck(vmpck)@m)
$\mkwd{\lor}$ ($\mkwd{\exists}$ #m. (#m < #l) $\mkwd{\land}$ CorruptImageOek(gvmId)@m)
$\mkwd{\lor}$ ($\mkwd{\exists}$ #m. (#m < #l) $\mkwd{\land}$ CorruptImageVmpck(maId1)@m)
$\mkwd{\lor}$ ($\mkwd{\exists}$ #m. (#m < #l) $\mkwd{\land}$ CorruptImageVmpck(maId2)@m)
\end{tcbTamarin2}
\caption{Strong integrity of attestation report}
\label{lem:attest}
\end{lemma}

\begin{table*}[t]
  \centering
  \scriptsize{} 
  \SetTblrInner{rowsep=0.6pt, colsep=3pt} 
  \begin{tblr}{
      colspec = {lX[c]c|X[l]cc|ccc}, 
      cell{odd[5-46]}{4-6} = {gray!20},
    }
    \toprule
    \toprule
    \addlinespace[2pt]
    \textsc{Properties} & \textsc{Description} & \# & \textsc{Lemma} & \textsc{Model} & \textsc{Steps} & \textsc{Runtime (min)} & \textsc{\#loc}\\
    \midrule
    \addlinespace[2pt]
    \textbf{Source} & \parbox{5.0cm}{Mitigating partial deconstructions.} & 1 & Types & \chmark & 938 & 1 & 40  \\
    \addlinespace[2pt]
    \hline
    \addlinespace[2pt]
    \textbf{Executability} & \parbox{5.0cm}{Functional correctness.} & 13 & Exe* & \chmark & - & 18 & 1360 \\

    \addlinespace[2pt]
    \hline
    \addlinespace[2pt]
    \textbf{Supporting} & \parbox{5.0cm}{Enforce termination and improve verification time} & 58 & Sup* & \chmark & - & 304 & 2600 \\

    \hline
    \SetCell[r=13]{m} \textbf{Secrecy} & \SetCell[r=13]{m} \parbox{5.0cm}{Secrecy of long-term keys and session keys. For example,  the secrecy of guest \vlst{VMPCK}, with the migration-enabled (\vlst{MA_EN=1}) policy (SecGvmVmpckIsSecret).} & \SetCell[r=13]{m} 13 &
    SecMaVmpckIsSecret & \chmark  & 6427 & \SetCell[r=13]{m} 107 & \SetCell[r=13]{m} 1000  \\
    \hline
    & & & SecGvmOekIsSecretMd & \chmark  & 1621 \\
    \hline
    & & & SecGvmOekIsSecret & \chmark  & 3455 \\
    \hline
    & & & SecGvmVmpckIsSecretMd & \chmark  & 1625 \\
    \hline
    & & & SecGvmVmpckIsSecret & \chmark  & 4436 \\
    \hline
    & & & SecMaVmrkIsSecret & \chmark  & 3418 \\
    \hline
    & & & SecFwVmrkIsSecret & \chmark  & 1619 \\
    \hline
    & & & SecKeyDerFromMaVmrkIsSecret & \chmark  & 12833 \\
    \hline
    & & & SecKeyDerFromFwVmrkIsSecret & \chmark  & 691 \\
    \hline
    & & & SecArkIsSecret & \chmark  & 2155 \\
    \hline
    & & & SecAskIsSecret & \chmark  & 2164 \\
    \hline
    & & & SecCekIsSecret & \chmark  & 2156 \\
    \hline
    & & & SecVcekIsSecret & \chmark  & 302 \\

    \hline
    \SetCell[r=18]{m} \textbf{Authentication} & \SetCell[r=18]{m} \parbox{5.0cm}{Authenticity of guest messages, exchanged between SNP-protected guest/migration agent and SNP firmware. These, for instance, specify that a message received by the firmware necessarily originated at the associated guest that was, at the time, executing on the same platform.} & \SetCell[r=18]{m} 18 &
    AuthFwGvmAgreeMsgAttestMd & \chmark  & 52 & \SetCell[r=18]{m} 22 & \SetCell[r=18]{m} 1600  \\
    \hline
    & & & AuthFwGvmWeakAgreeMsgAttest & \chmark  & 282 \\
    \hline
    & & & AuthFwGvmAgreeMsgAttest & \xmark  & 27 \\
    \hline
    & & & AuthFwGvmAgreeMsgKeyDerMd & \chmark  & 47 \\
    \hline
    & & & AuthFwGvmWeakAgreeMsgKeyDer & \chmark  & 282 \\
    \hline
    & & & AuthFwGvmAgreeMsgKeyDer & \xmark  & 27 \\
    \hline
    & & & AuthGvmFwAgreeMsgAttestMd & \chmark  & 129 \\
    \hline
    & & & AuthGvmFwWeakAgreeMsgAttest & \chmark  & 56 \\
    \hline
    & & & AuthGvmFwAgreeMsgAttest & \xmark  & 29 \\
    \hline
    & & & AuthGvmFwAgreeMsgKeyDerMd & \chmark  & 129 \\
    \hline
    & & & AuthGvmFwWeakAgreeMsgKeyDer & \chmark  & 56 \\
    \hline
    & & & AuthGvmFwAgreeMsgKeyDer & \xmark  & 29 \\
    \hline
    & & & AuthFwMaAgreeMsgVmrk & \chmark  & 7 \\
    \hline
    & & & AuthFwMaAgreeMsgExport & \chmark  & 5 \\
    \hline
    & & & AuthFwMaAgreeMsgImport & \chmark  & 5 \\
    \hline
    & & & AuthMaFwAgreeMsgVmrk & \chmark  & 6 \\
    \hline
    & & & AuthMaFwAgreeMsgExport & \chmark  & 10 \\
    \hline
    & & & AuthMaFwAgreeMsgImport & \chmark  & 6 \\

    \hline
    \SetCell[r=5]{m} \textbf{Freshness} & \SetCell[r=5]{m} \parbox{5.0cm}{Uniqueness of certain events. For instance, whether each message of guest employes a fresh counter (FreshNoMsgRevealFromNonceReuse). } & \SetCell[r=5]{m} 138 &
    FreshNoMsgRevealFromNonceReuse & \chmark & 4096 & \SetCell[r=5]{m} 138 & \SetCell[r=5]{m} 480  \\
    \hline
    & & & FreshMaVmrkInstallIsUnique & \chmark  & 20 \\
    \hline
    & & & FreshAttReportFreshness & \chmark  & 10 \\
    \hline
    & & & FreshKeyDerFromFwVmrkIsGvmUnique & \chmark  & 177 \\
    \hline
    & & & FreshKeyDerFromMaVmrkIsGvmUnique & \chmark  & 17825 \\

    \hline
    \SetCell[r=6]{m} \textbf{Attestation} & \SetCell[r=6]{m} \parbox{5.0cm}{Attestation report integrity and authenticity. For instance, if an attestation report states that a guest with a certain configuration is executing on a particular platform, then this must indeed be true for that guest. } & \SetCell[r=6]{m} 6 &
    AttReportAuthenticityMd & \chmark  & 505 & \SetCell[r=6]{m} 47 & \SetCell[r=6]{m} 900  \\
    \hline
    & & & AttReportAuthenticity & \chmark  & 16 \\
    \hline
    & & & AttReportIntegrityMd & \chmark  & 554 \\
    \hline
    & & & AttReportWeakIntegrity & \chmark  & 2028 \\
    \hline
    & & & AttReportStrongIntegrity & \xmark  & 52 \\
    \hline
    & & & AttBackAndForthMig & \xmark & 71 \\
    \hline
    & \parbox{3.5cm}{\raggedleft$\Sigma$}  & 114 &  & & & 365 & 7845 \\
    \bottomrule
    \bottomrule
  \end{tblr}
  \caption{A summary of the analyzed properties. All properties are
    automatically verified (\chmark{}) or falsified (\xmark{}). We leverage oracle rankings
    based on the "smart" heuristic, with supporting lemmas, for proof
    guidance. The suffix ``Md'' indicates that a guest was launched with the
    migration-disabled policy. \textsc{steps} denote proof length. The final
    lemma holds if the \tlst{-DENFORCE_MIGRATION_POLICY} flag is enabled. }
  \label{table:propsummary}
\end{table*}

\subsection{Positive Results}

On a positive note, all secrecy and freshness properties, and the majority of
authentication and attestation properties, have been successfully proven. 
Hence, our results show that the SEV-SNP software interface does indeed provide almost all the desired security guarantees, while exhibiting some potential drawbacks discussed in the following subsections.

The positive results hold in a very general setting with an unbounded number of platforms, VMs, guest owners and sessions. Furthermore, they hold assuming an adversary capable of corrupting arbitrary agents.
The verification was enabled by supporting lemmas and proof
strategies based on the oracle rankings.
Both of them were utilized to either enforce termination or decrease verification time.
Finally, many security properties
were specified in a very granular manner, which was instrumental in
proving other security properties.

A substantial amount of work went into proving the lemma
\tlst{FreshNoMsgRevealFromNonceReuse}; the majority of the supporting lemmas were
specifically leveraged for this purpose. This is mainly because we allow the \snpgvm{} state to be swapped out and swapped in an unbounded number of times.
Consequently, we had to prove that the \snpfw{} and \snpgvm{} message counters are
synchronised with each swap operation, i.e., they are either equal or
the \snpfw{} counter exceeds the \snpgvm{} counter by two.  The corresponding
lemmas are prefixed with \tlst{SupFWAndGVMMsgCountersAreInSync}.








\subsection{Platform Confusion Attacks}
We identified five \emph{formal attacks} in our model; four on authentication properties and one on an attestation property. 
In all the attacks, the platform-agnostic nature of guest messages, along with the migration feature, is used by the adversary to guide the system into a state we believe is undesirable.

The core issue is that guest messages lack binding to a specific platform. We believe that this is a design decision made to facilitate the seamless migration of guests. However, this allows guest messages sent from one platform to be received on another, although it seems reasonable to expect, e.g., that guests only accept messages sent by the firmware on the same platform.


We would like to emphasize that we have not attempted to launch these attacks
in practice---this would require the development of a malicious hypervisor and a kernel driver for SNP-protected guests, both of which support AMD SEV-SNP live migration. Note that no such open-source hypervisor exists (e.g., AMD has still not added support for migration of SNP-protected guests to their QEMU fork~\cite{amdqemufork}).

Similarly, we do not claim that these attacks have a direct impact on security of the deployed systems, but we do discuss potential impact as well as mitigation strategies in the following section.
Finally, we note that we did not anticipate these properties to fail---they were discovered purely by formal analysis.



In the four authentication attacks, a message originating on one platform is accepted on a different platform. The four attacks are brought about by all possible combinations of the following two binary parameters: (1) where the message originated (at \snpfw{} and is intended for \snpgvm{}, at \snpgvm{} and is intended for \snpfw{}) and (2) the type of request being sent or responded to (attestation report request, key derivation request). Note that minor variants exist; for example, the \snpgvm{} can be migrated before or after the \snpfw{} processes the request.

\begin{figure*}
  \centering
  \vspace{-15pt}
  \includegraphics[width=14.5cm]{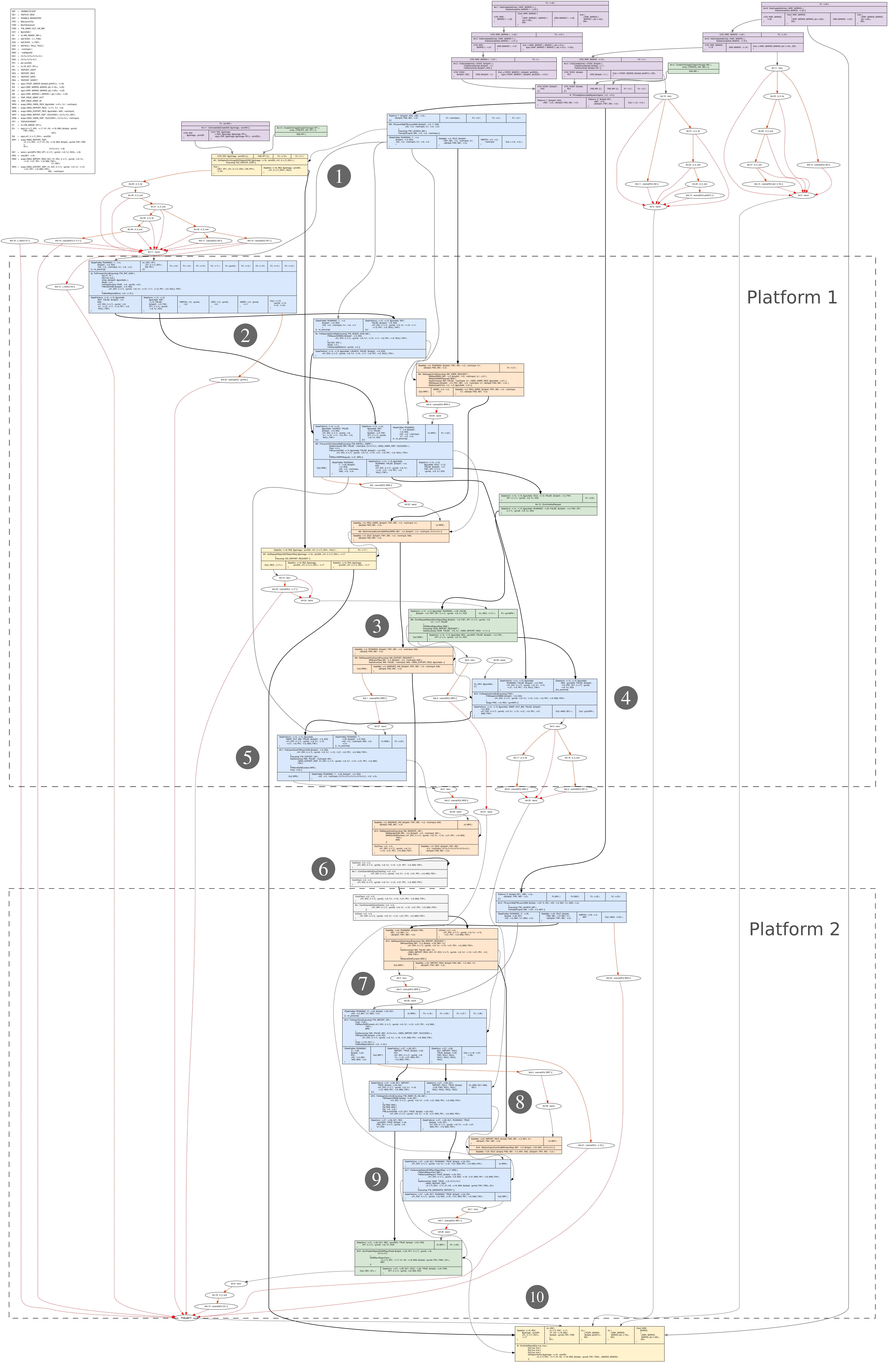}
  \caption[]{An attack trace for the AttReportStrongIntegrity lemma:
\begin{enumerate*}[label=\protect\circled{\arabic*}]
\item \snpgo{} deploys a guest image and policy;
\item \snpfw{} launches guest on the first platform;
\item\label{trace:reportreq}\snpgvm{} requests an attestation report (\tlst{MSG_REPORT_REQ});
\item \snpfw{} immediately swaps out the guest;
\item \snpfw{} exports the guest context;
\item \snpma{} transfers the guest context to \textsc{ma'} on the second platform;
\item \textsc{fw'} imports the guest context;
\item \textsc{fw'} swaps in the guest;
\item \textsc{fw'} receives \tlst{MSG_REPORT_REQ} from~\ref{trace:reportreq}, and issues an attestation report;
\item \snpgo{} verifies the report.
\end{enumerate*}
In the end, \snpgo{} erroneously believes that the
attestation report was requested on the second platform (where it was generated), whereas the \snpgvm{} was
executing on the first platform at that time.
}
\label{fig:attacktrace}
\end{figure*}

The steps to reproduce the attacks wherein the \snpgvm{} sends a request on one platform, and the adversary forwards it to the \snpfw{} on another platform can be seen in~\Cref{fig:attacktrace}. The \snpfw{} on the destination platform
incorrectly believes that the \snpgvm{} sent the message while it was executing
on the destination platform, whereas in reality the \snpgvm{} was
executing on the source platform at the time.
Because the request could have been for either an attestation report or for key derivation, the described attack shows that neither of the \tlst{AuthFwGvmAgreeMsgAttest} (see Lemma~\ref{lem:auth}) and \tlst{AuthFwGvmAgreeMsgKeyDer} properties holds. In both cases there is no agreement on
the value of \tlst{chipId}. As a technical sidenote, to find an attack trace faster in the final model, we added several preconditions to the lemmas that restricted the set of considered traces.

In the remaining two authentication attacks, the \snpfw{} sends a response to the corresponding guest request, and the adversary migrates the \snpgvm{} to another platform (where the \snpgvm{} is resumed and waits for a response) before forwarding the response to the \snpgvm{}.

However, if the adversary now forwards the response from the
\snpfw{} on the source platform to the \snpgvm{}, the \snpgvm{} will accept it. The described attack falsifies the \tlst{AuthGvmFwAgreeMsgAttest}
and \tlst{AuthGvmFwAgreeMsgKeyDer} properties.

Similarly, in the attack on the violated attestation property, the attestation report that the \snpgo{} verifies was not generated on the platform where the \snpgvm{} requested the attestation report. Instead, it was generated on the platform that the \snpgvm{} was migrated to.
As a consequence, \tlst{AttReportStrongIntegrity} (see Lemma~\ref{lem:attest}) is falsified because \tlst{chipId} can differ from the anticipated value. The corresponding trace is shown in~\Cref{fig:attacktrace}.

When considering changes to the model that would completely prevent the platform confusion attacks, we decided against changes that appear to translate poorly to practice.
In particular, we have considered to make the guest messages platform-specific by employing a fresh \vlst{VMPCK} after every migration.
While this change to the model would address the failing formal properties, we are reluctant to recommend it as a practical solution. Namely, it would require modifications to guest pages upon each guest import, and it is unclear to us how to do this while preserving seamless migration, i.e., without having to reboot the guest.

In spite of that, we show that the scope of discovered formal attacks is limited by stating \emph{weak} versions of all failing lemmas and proving that they all hold.
Using the attestation integrity property as an example, the weak variant says that if a \snpgo{} verifies the attestation report, then it must have been requested by a \snpgvm{} running \emph{on some platform}---the \tlst{chipId} value in the attestation report does not have to match the \tlst{chipId} value of the platform where the \snpgvm{} was running at the time of the request. Weaker variants of authentication properties are similarly defined.


\subsection{Back-and-forth migration attack}
The formal attacks in the previous section show that if a guest is configured to allow migration, the \vlst{CHIP_ID} and \vlst{COMMITTED_TCB} values in the verified attestation report do not need to correspond to the platform where the guest was running, at or near the time it requested the attestation. Hence, a malicious cloud provider may execute the described attacks and successfully deceive a guest owner into believing that a guest is executing critical functionality on an up-to-date platform when, in reality, it is not.

As a matter of fact, the issue just described persists independently of the discovered platform confusion attacks. Even if we addressed the attacks somehow, a malicious cloud provider might still be able to migrate a guest just before the guest requests an attestation report. Specifically, the cloud provider could run the guest on a vulnerable platform, but obtain attestation reports by briefly migrating the guest to the secure platform, thus creating an illusion that the guest is consistently managed by secure, up-to-date firmware.

The property that is violated, \tlst{AttNoBackAndForthAttack}, considers the
guest's execution history. It states that
not only was the guest executing on a platform whose firmware TCB version
matches the version in the attestation report, but also that every previous
guest launch occurred on firmware with the same TCB version.
If the version in the attestation report reflects an up-to-date firmware image, a
third party can be confident that the guest has consistently operated on a
secure platform.

Although in practice we want to ensure that the TCB version never "decreases" across
platforms, here for simplicity we focus on maintaining version equality.
To automatically find an attack trace, we utilize the
\tlst{AttBackAndForthMigAttack} lemma.

The most direct mitigation for the above issues, including the just described
\emph{back-and-forth migration attack}, is to enforce a TCB policy and other
migration policies in the migration agent.
In fact, the SEV-SNP firmware ABI specification~\cite{sevsnp:spec} already states the following:
\begin{shadedquotation}
The MA is responsible for supplying the \vlst{VMRK} during the launch process and for enforcing the guest
migration policy.
\end{shadedquotation}
However, it does not give any details on what the migration policy might be. Hence, we propose that the SEV-SNP ABI specification be updated to:
\begin{itemize}
\item make explicit the migration policy that a compliant migration agent should enforce, and
\item clarify the (lack of) guarantees on the \vlst{CHIP_ID} and \vlst{COMMITTED_TCB} values
in the verified attestation report when migration is enabled, and their dependence on the
migration agent behavior.
\end{itemize}

The \tlst{-DENFORCE_MIGRATION_POLICY} flag enables the policy in our model.
It adds a simple equality check to ensure that the firmware images on both the
source and destination platforms have the same TCB version.

\subsection{Post-analysis insights and lessons learned}
After performing the formal analysis of SEV-SNP, several key insights and lessons emerged.

\emph{There is a tension between the simplicity of the SEV-SNP software interface and the security guarantees of the system whose critical components are underspecified.}

While SEV-SNP is richer in terms of both usability features and security guarantees compared to both previous iterations, namely SEV and SEV-ES, the software interface of SEV-SNP is also more streamlined in some ways. First, the launch procedure has been simplified as the launch protocol no longer exists.
Second, the mechanism of migration is now outsourced to a new entity called a Migration Agent. Third, the key schedule is simpler as there are fewer key dependencies. This includes the absence of, for instance, the \vlst{TIK}, \vlst{TEK}, \vlst{KEK} and \vlst{KIK}. Fourth, no Diffie-Hellman operations, which are typically costly to analyze, are necessitated by the specification.

Although these changes make the SEV-SNP software interface more amenable to formal analysis, the added simplicity is a double-edged sword. Critical parts of the system, such as the migration agent, are left underspecified.
Our formal analysis demonstrates that it is essential to have clear guidelines for building migration agents, or a canonical migration agent implementation, so as to ensure critical security properties.

\emph{There is a tension between the ability to easily and transparently migrate an SNP-secured guest and the guarantees provided by the attestation protocol to the relying party.}

Many cloud workloads of today support live migration without perceived downtime.
However, our formal analysis shows that the platform-identifying data in the attestation reports cannot be trusted in all circumstances. 
It would be beneficial if the attestation report included information about
a set of platforms on which the guest can execute. The guest owner does not
need to know or care on which particular machine it is currently executing
because that information may become deprecated soon (e.g., as soon as the guest migrates to another machine).

\emph{Multiset rewriting and the Tamarin prover are suitable tools for the formal analysis of complex software interfaces; however, SEV-SNP pushes the boundaries of these methods, requiring further theoretical advancements to enable routine, full-fidelity analysis of such systems.}

While the final, fully automated formal analysis can be conducted in eight hours, this is the result of much effort to specify the supporting lemmas, implement the proof-guiding oracles and to simplify the model without sacrificing fidelity. 
We believe theoretical advancements in the area of automated verification that would enable the compositionality and reuse of components and proofs would be immensely practical for the analysis of similar systems. For example, a significant effort was dedicated to verifying the freshness of nonces used in the GCM encryption, where the supporting scaffolding substantially increased the overall complexity of the analysis.

We envision a compositional approach that would allow for the analysis of a simplified version of the system, once nonce reuse has been ruled out definitively. However, no automated tools support such reasoning at the moment.


\emph{The novel software interface differs too significantly from that of SEV(-ES) to allow the results to be carried over.}

As a matter of fact, even reusing parts of our model to analyze the security of the SEV and SEV-ES software interfaces is currently infeasible. Namely, the described differences are simply too vast; for instance, the guest messages, which our analysis focuses on, were not supported previously.
However, we believe our model is sufficiently modular to be easily extended and used
to analyze additional aspects of the SEV-SNP software interface, even within an
unbounded number of migrations.  In addition, it provides a useful reference for
any similar modelling and verification endeavor, especially when the tools used
are based on multiset rewriting.







\section{Related Work}\label{sec:relatedwork}




There is a vast body of literature devoted to AMD SEV technologies in general.
However, to the best of our knowledge, there has been no attempt to build a
symbolic model and analyze the security of SEV-SNP, including
its ABI.
Apart from a work by Antonino et al.~\cite{AntoninoDW23} that
considers
SGX and pre-SNP SEV software interface
to propose and formally verify
a flexible remote attestation protocol,
most of the existing SEV research focuses on
its implementation. Although we find this line of work orthogonal to ours,
here we provide a short overview.

Attacks on SEV and SEV-ES exploit nested page
tables~\cite{10.1145/3193111.3193112}, external interfaces such as
\vlst{virtio} devices~\cite{10.1145/3433667.3433668} and direct memory access~\cite{Li1},
the signature verification mechanism of the AMD-SP OS~\cite{10.1145/3319535.3354216},
the lack of memory integrity protection~\cite{Wilke2020SEVurityNS},
TLB entries~\cite{Li4}, block permutation-agnostic measurements of VM images~\cite{9474294},
guest address space identifiers~\cite{Li3}, and
the power-reporting interface~\cite{10.1007/978-3-031-35504-2_3}.
Attacks on SEV-SNP target \vlst{#VC}~\cite{wesee} and \vlst{0x80} interrupts~\cite{heckler},
use software-based fault injection~\cite{Buhren, Zhang2023Cachewarp}, and
rely on ciphertext side channels~\cite{Li2, Li5}.
Additionally, Google has identified multiple issues with SEV-SNP firmware~\cite{projectzerosevsnp}.

Other works have used formal techniques to reason about the interfaces of other
trusted execution environments~\cite{10.1007/978-3-030-61078-4_8,
10.1145/2810103.2813608, 274711, 7962001, 10.1145/3133956.3134098, 10221878} and trusted platform modules
(TPMs)~\cite{10.1007/978-3-642-19751-2_8, 10.1145/2714576.2714610}.  Notably,
attestation
schemes have received a lot of attention: Intel SGX Data Center Attestation
Primitives~\cite{10.1007/978-3-030-63406-3_16} and TDX~\cite{9448036}, ARM CCA~\cite{10373038}, and ECC-based~\cite{10.1145/3320269.3372197} and TPM
2.0-based~\cite{10.1145/3320269.3372197} Direct Anonymous Attestation schemes.
Moreover, CCA is the first VM-based TEE architecture
with a formally verified firmware~\cite{280904}.

\section{Conclusion}\label{sec:conclusion}

In this paper, we developed the first, comprehensive formal model of the AMD
SEV-SNP software interface. The model covers the guest launch, remote
attestation, key derivation, page swap and live migration features.  We produced
automated formal proofs for the most important secrecy, authenticity,
attestation, and freshness properties, including the proof of correct stream
cipher usage.

While most of the relevant security properties were successfully verified, our analysis shows that several authentication and attestation properties do not hold
when SNP-protected guests are launched with migration-enabled policies
due to a lack of platform binding in guest messages.
Despite identifying these formal attacks, there is still work to be done; most pressing are a practical confirmation of the attacks, establishing the severity of their consequences and, if there are any, designing and implementing mitigations that preserve seamless migration.








\section*{Acknowledgments}
This work has been supported by the European Union through the European Regional
Development Fund, under the grant KK.01.1.1.01.0009 (DATACROSS), and the project AutoDataLog, a cooperation between the Faculty of Science and AVL-AST d.o.o.\ Croatia.
We would also like to thank University Computing Center of Zagreb (SRCE) for providing us with the computing resources.

\nocite{*}
{
  \footnotesize
  \bibliographystyle{./ieee-bibliography-template/IEEEtran}
  \bibliography{./ieee-bibliography-template/IEEEabrv, amd-sev-snp}
}

\begin{IEEEbiography}[{\includegraphics[width=1in,height=1.25in,clip,keepaspectratio]{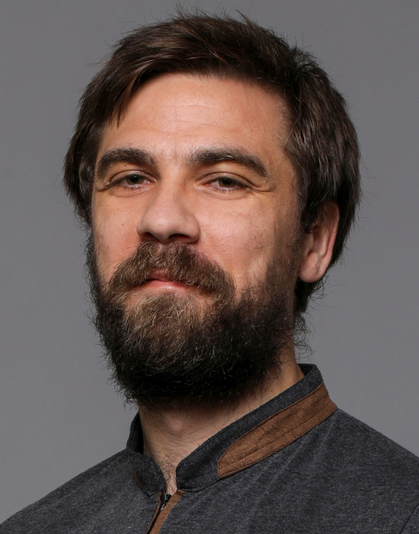}}]{Petar Parad\v{z}ik}
received an MSc in Computer Science and Mathematics from
the Department of Mathematics, Faculty of Science, University of Zagreb. He is
currently a PhD student and teaching assistant at the Faculty of Electrical
Engineering and Computing in Zagreb. His research interests include formal
methods, automated verification, and applied cryptography.

\end{IEEEbiography}

\begin{IEEEbiography}[{\includegraphics[width=1in,height=1.25in,clip,keepaspectratio]{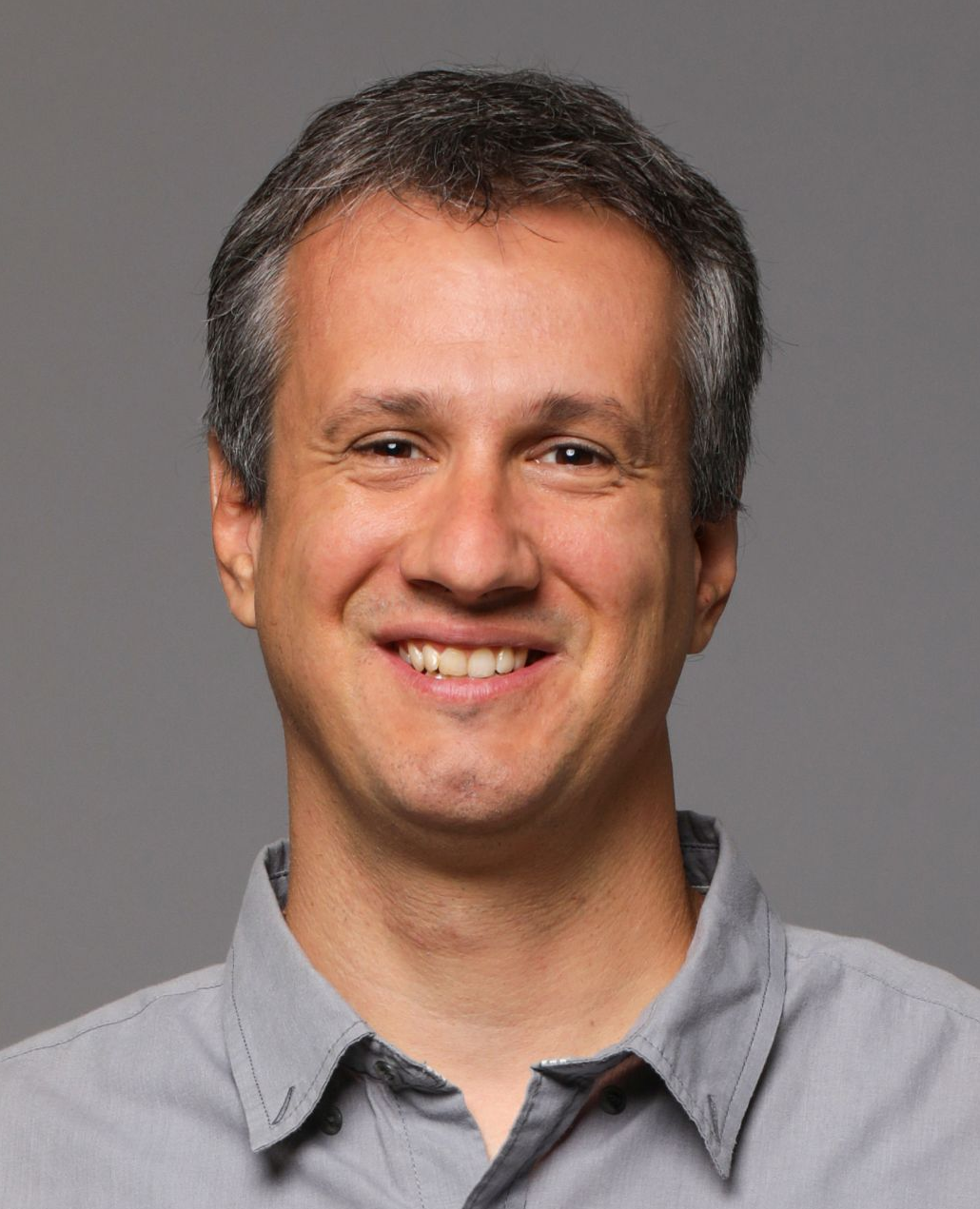}}]{Ante Derek}
(Member, IEEE) is currently an
Associate Professor with the Faculty of Electrical Engineering and Computing,
University of Zagreb. He participates in a number of national and EU-funded
projects in the area of computer security. His research interests include the
area of applying formal methods to problems in computer security, privacy, and
cryptography.

\end{IEEEbiography}

\begin{IEEEbiography}[{\includegraphics[width=1in,height=1.25in,clip,keepaspectratio]{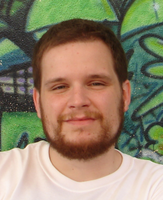}}]{Marko Horvat}
received a DPhil in Computer Science from the University of Oxford,
UK. He is currently working as Assistant Professor at the Department of
Mathematics, Faculty of Science, University of Zagreb, Croatia. His research
interests range from formal verification of security protocols to computable
analysis and topology.
\end{IEEEbiography}

\vfill

\end{document}